\documentclass[final,onefignum,onetabnum]{siamart220329}
\usepackage{braket,amsfonts}

\usepackage{array}

\usepackage[caption=false]{subfig}
\usepackage{pgfplots}

\newsiamthm{claim}{Claim}
\newsiamremark{remark}{Remark}
\newsiamremark{hypothesis}{Hypothesis}
\crefname{hypothesis}{Hypothesis}{Hypotheses}

\usepackage{algorithmic}
\usepackage{hyperref}
\usepackage{graphicx,epstopdf}

\Crefname{ALC@unique}{Line}{Lines}

\usepackage{amsopn}

\usepackage{xspace}
\usepackage{bold-extra}
\usepackage[most]{tcolorbox}

\colorlet{texcscolor}{blue!50!black}
\colorlet{texemcolor}{red!70!black}
\colorlet{texpreamble}{red!70!black}
\colorlet{codebackground}{black!25!white!25}

\lstdefinestyle{siamlatex}{%
  style=tcblatex,
  texcsstyle=*\color{texcscolor},
  texcsstyle=[2]\color{texemcolor},
  keywordstyle=[2]\color{texemcolor},
  moretexcs={cref,Cref,maketitle,mathcal,text,headers,email,url},
}

\tcbset{%
  colframe=black!75!white!75,
  coltitle=white,
  colback=codebackground, 
  colbacklower=white, 
  fonttitle=\bfseries,
  arc=0pt,outer arc=0pt,
  top=1pt,bottom=1pt,left=1mm,right=1mm,middle=1mm,boxsep=1mm,
  leftrule=0.3mm,rightrule=0.3mm,toprule=0.3mm,bottomrule=0.3mm,
  listing options={style=siamlatex}
}

\newtcblisting[use counter=example]{example}[2][]{%
  title={Example~\thetcbcounter: #2},#1}

\newtcbinputlisting[use counter=example]{\examplefile}[3][]{%
  title={Example~\thetcbcounter: #2},listing file={#3},#1}

\DeclareTotalTCBox{\code}{ v O{} }
{ 
  fontupper=\ttfamily\color{black},
  nobeforeafter,
  tcbox raise base,
  colback=codebackground,colframe=white,
  top=0pt,bottom=0pt,left=0mm,right=0mm,
  leftrule=0pt,rightrule=0pt,toprule=0mm,bottomrule=0mm,
  boxsep=0.5mm,
  #2}{#1}

\patchcmd\newpage{\vfil}{}{}{}
\title{Splitting Alternating Algorithms for Sparse Solutions of Linear Systems with Concatenated Orthogonal Matrices\thanks{Submitted to the editors November 2, 2024.
\funding{This work was supported by the National Key Research and Development Program of China (2023YFA1009302), National Natural Science Foundation of China (12471295, 12426306, 12071307, 12371305), Guangdong Basic and Applied Basic Research Foundation (2024A1515011566), and the Hetao Shenzhen-Hong Kong Science and Technology Innovation Cooperation Zone Project (HZQSWS-KCCYB-2024016).}}}

\author{Yun-Bin Zhao\thanks{Shenzhen International Center for Industrial and Applied Mathematics, SRIBD, The Chinese University of Hong Kong, Shenzhen, China (\email{yunbinzhao@cuhk.edu.cn}). }
\and Zhong-Feng Sun\thanks{School of Mathematics and Statistics, Shandong
University of Technology, Zibo, Shandong,  China (\email{zfsun@sdut.edu.cn}) } }

\headers{Splitting Alternating Algorithms}{Yun-Bin Zhao and Zhong-Feng Sun}

\newtheorem{Thm}{Theorem}[section]

\newtheorem{Lem}[Thm]{Lemma}

\newtheorem{Rem}[Thm]{Remark}

\begin{document}
\maketitle


\begin{abstract}
 A class of splitting alternating algorithms is proposed for finding the sparse solution of linear systems with concatenated orthogonal matrices. Depending on the number of matrices concatenated, the proposed algorithms are classified into the two-block splitting alternating algorithm (TSAA) and the multi-block splitting alternating algorithm (MSAA). These algorithms aim to decompose a large-scale linear system into two or more coupled subsystems, each significantly smaller than the original system, and then combine the solutions of these subsystems to produce the sparse solution of the original system. The proposed algorithms only involve  matrix-vector products and reduced orthogonal projections. It turns out that the proposed algorithms are globally convergent to the sparse solution of a linear system if the matrix (along with the sparsity level of the solution) satisfies a coherence-type condition. Numerical experiments indicate that the proposed algorithms are very promising and can quickly and accurately locate the sparse solution of a linear system with significantly fewer iterations than several mainstream iterative methods.
\end{abstract}



\section{Introduction}
 The sparse solution of an underdetermined linear system has gained significant interest in the field of science and engineering \cite{BDE09, E10,FR13,Z18,AH21,LW21}. Many practical problems can be formulated as finding the sparse solution to a linear system, which commonly arises in compressive sampling \cite{C06, D06}, signal and image reconstruction \cite{E10,M16,MEH19,AH21}, linear inverse problem \cite{BT09,S15,SZZ23,STY24}, model selection \cite{HTW15,GJ23}, and wireless network \cite{CSDRK17,LCG19}.  The fundamental model for the sparse solution of a linear system  can be cast as
  \begin{equation} \label{Sparse-Sol-LS} \min\{\|x\|_0: ~ y= Ax \}, \end{equation}
where $ y \in \mathbb{R}^m$ is a given vector,  $A$ is a given $m \times n $ matrix with $ m \ll n, $ and $ \|x\|_0$  counts the number of nonzero entries of $x\in \mathbb{R}^n. $ The data pair $(y,A)$  may have different interpretations depending on application context. For instance, in compressive sampling \cite{C06,D06}, $A$ is referred to as the measurement matrix and the entries of $y$ are the acquired measurements of the signal; in wireless communication \cite{CSDRK17,LCG19},  $A$ is called the channel-gain matrix, and the entries of $y$ are the received signals at the terminals. Before proceeding, let us review some existing algorithms developed over the past decades for solving (\ref{Sparse-Sol-LS}).

(i) \emph{Convex optimization methods}. Problem (\ref{Sparse-Sol-LS}) can often be well approximated  by the $\ell_1$-minimization model $\min \{ \|x\|_1: y=Ax\} $ which is a convex optimization model (see, e.g., \cite{CDS98,CT05}).
The Lasso model $\min \{\|y-Ax\|_2^2 + \lambda\|x\|_1 \}, $ where $ \lambda $ is a positive parameter, is also a popular convex approximation  of (\ref{Sparse-Sol-LS}) \cite{BPCPE10,AR14,HTW15}. Replacing $\|x\|_0 $ in  (\ref{Sparse-Sol-LS}) with a nonconvex `merit function for sparsity' and employing a linearization technique, one can develop the  reweighted $\ell_1$-minimization methods for this problem \cite{CWB08,ZL12,Z18}.
Additionally, the so-called dual-density-based approach in \cite{ZL17, Z18} is also a convex optimization method for this problem. Convex optimization problems are typically solved using interior-point methods which are efficient for reasonably sized problems but have a significantly higher computational cost compared to the next two classes of methods.

(ii) \emph{OMP-type methods.}  The orthogonal matching pursuit (OMP) was introduced in the statistics literature several decades ago and later extended to signal processing \cite{PRK93,MZ93,TG07,E10}. Although OMP performs well for (\ref{Sparse-Sol-LS}) in many cases, it is relatively slow in locating the high-sparsity-level solution of large-scale problems. This is because the total number of iterations required by OMP is at least equal to the number of nonzero entries in the solution. This issue also affects the variants of OMP, including the stagewise orthogonal matching pursuit \cite{DTDS12} and weak orthogonal matching pursuit \cite{T00}.

(iii) \emph{Thresholding-type methods.}   The iterative hard thresholding (IHT) \cite{BD08} and hard thresholding pursuit (HTP) \cite{F11} are also popular methods for solving (\ref{Sparse-Sol-LS}). Combining matching pursuit \cite{MZ93} with hard thresholding gives rise to the compressive sampling matching pursuit (CoSaMP) \cite{NT09} and subspace pursuit (SP) \cite{DM09}. Soft-thresholding methods, including the iterative shrinkage-thresholding algorithms (ISTA) and fast iterative shrinkage-thresholding algorithm (FISTA) \cite{BT09}, are also widely used for finding sparse solutions of linear systems. However, it is known that IHT, ISTA, and FISTA converge slowly and their performance is sensitive to the choice of parameters such as stepsize or soft-thresholding parameter. Recent developments in thresholding-type methods can be found in \cite{SBRJ19, FS19, Z20, MZ20, GLMNQ23, SZZ23, HSWW24, K24, EYNS24}.

Problem (\ref{Sparse-Sol-LS}) is NP-hard \cite{N95} and developing an efficient general-purpose algorithm for it is challenging. The aim of this paper is to develop an efficient, customized algorithm for a class of structured large-scale problems involving matrices that are concatenations of orthogonal matrices:
\begin{equation} \label{2A2A} A= [\Phi_1, \Phi_2, \cdots, \Phi_p], \end{equation}
where $ \Phi_i ,$ $ i=1, \dots, p$  are  $m\times m$ orthogonal matrices. Such matrices have long been used as measurement matrices in signal and image recovery \cite{DH01,EB02,E10,FN02,GN03}, and they appear in applications like distributed compressed sensing and distributed sensor networks \cite{KVC12,PEK14,WZWCY16}. Concatenated orthogonal matrices have both theoretical and practical advantages due to their reduced coherence, which is crucial for robust signal recovery \cite{E10,GN03,KVC12}.
 However, the existing methods such as $\ell_1$-minimization, OMP and thresholding methods use the matrix as a whole, failing to exploit its internal structure. Compressive sampling theory \cite{CT05,E10,FR13} indicates that the performance of a signal-reconstruction algorithm is  closely tied to the structure of the measurement matrix  and its properties such as reduced coherence \cite{E10}, restricted isometry property \cite{CT05}, null space property\cite{FR13}, and range space property  \cite{Z18}. Both simulations and practical applications demonstrate that the matrix structure significantly impacts the performance of signal processing algorithms. Therefore, it is essential to develop an algorithm that fully leverages the matrix structure while maintaining low computational cost for solving large-scale problems.

In this paper, we develop the splitting alternating algorithms (SAA) for finding sparse solutions to large-scale linear systems with concatenated orthogonal matrices. When two matrices are concatenated, the algorithm is called the two-block splitting alternating algorithm (TSAA); for more than two matrices, it is referred to as the multi-block splitting alternating algorithm (MSAA). Our algorithms aim to solve the problem at a low computational cost by fully utilizing the matrix structure in (\ref{2A2A}).  Starting from an initial partition of $y,$ say $ y= y_1+ \cdots + y_p$, we split the original large-scale linear system into $p$  coupled subsystems  $ y_i = \Phi_i x_i,$ $i= 1, \dots, p,$   solve these subsystems, and update the current partition of $y$ alternately.  The idea of alternating iteration has been widely used in the alternating direction method of multipliers (ADMM) and distributed optimization approaches \cite{BPCPE10,GOY16, ZL22REV,YGJYXS22}. SAA can also be viewed as a decomposition method for large-scale sparse optimization problems. We prove that the proposed algorithms globally converge to the sparsest solution of a linear system if the mutual coherence of the matrix and the sparsity level of the solution jointly meet a certain condition. Numerical results indicate that the proposed algorithms can often solve a wide range of problems (\ref{Sparse-Sol-LS}) with just a few iterations. Moreover, the overall performance (success rate, convergence speed, and robustness in signal and image reconstruction) of the proposed algorithms is superior to many existing methods.
 
The paper is organized as follows. Section 2 describes the two-block splitting alternating algorithm, and its global convergence is shown in Section 3. The multi-block algorithm is presented in Section 4, and its global convergence is shown in Section 5. Finally, numerical results are given in Section 6.

\section{Two-block splitting alternating algorithms}
In this section, we describe the algorithm in two-block cases, while the  multiple-block cases will be studied in later sections. Let us first introduce some notations.

\subsection{Notation} For $ u, v \in \mathbb{R}^m$, we simply write the vector  $ x =  \left[\begin{array}{c}
  u \\
  v \\
  \end{array}\right] \in \mathbb{R}^{2m} $ as $ x= (u,v) $ when no ambiguity arises.  $x\in \mathbb{R}^n$ is said to be $K$-sparse if it has at most $K$ nonzero entries. We  denote the support of $x  $ by $\textrm{supp}(x)=\{i : x_i\neq 0\}.$   Given $S\subseteq \{1,\ldots,n\},$   the  cardinality and complement set of $S$ are denoted by $|S|$  and  $\overline{S} =\{1,\ldots,n\}\setminus S,$ respectively.   The vector $x_{S} $ is obtained from $x\in \mathbb{R}^n$ by retaining the entries   indexed by $S$ and   zeroing out (or simply removing) those entries indexed by $\overline{S}$. By zeroing out entries, $ x_S$ remains $n$-dimensional, while by removing entries,  $ x_S \in \mathbb{R}^{|S|} . $ The dimension of $ x_S$ is clear from the content. Sorting the absolute values of the entries of $x\in \mathbb{R}^n $ in  descending order $|x_{i_1}| \geq |x_{i_2}| \geq \dots \geq |x_{i_n}|, $ we denote the index set of the  $K$ largest absolute entries of $x  $ by $\mathcal{L}_K(x) =\{i_1, i_2, \dots, i_K\} .$   When two indices are tied, choose the smaller one.  The hard thresholding operator $\mathcal{H}_K(x)$  retains the  $K$ largest entries in magnitude and zeroes out other entries, so $\mathcal{H}_K(x) = x_S $ with $S={\mathcal{L}_K(x)}. $
  Given $S\subseteq \{1,\ldots,n\}$ and a matrix  $A =[a_1, \dots, a_n]\in \mathbb{R}^{m\times n} $ with columns $ a_i$, we denote by $A_S$ the submatrix of $A$ consisting of the columns indexed by $S.$ We  denote  the mutual coherence of $A$ by $ \label{mutcoh} \mu(A) = \max_{i \not= j} \frac{|a_i^T a_j|}{\|a_i\|_2\|a_j\|_2} $, which represents the largest absolute inner product between the normalized columns of $A.$

\subsection{Algorithm description}
Consider the matrix $A=[\Phi_a, \Phi_b],$  where   $\Phi_a, \Phi_b \in \mathbb{R}^{m\times m}  $ are orthogonal.  Suppose that $ x^*= (x_a^*, x_b^*)\in \mathbb{R}^{2m} $ is the   $K$-sparse solution to the system $ y= Ax$, where   $x_a^*, x_b^*\in \mathbb{R}^{m}.$ Then $y$ can be written as    $ y= y_a^*+ y_b^*$ with
$ y_a^*= \Phi_a x_a^* $ and $ y_b^* =  \Phi_b x_b^* .$ Such a partition of $y,$ determined by $x^*,$ is referred to as the {\it optimal partition} of $y.$
To find the solution $x^*, $  it is sufficient to find the optimal partition $(y_a^*, y_b^*)$ of $y.$    This motivates one to develop a method that can iteratively search for the optimal partition of $y.$  Let $( x_a^{(k)}, x_b^{(k)}) $ be the current iterate which is $K$-sparse. We adopt the following alternating search and partition strategy of $y$ to generate certain intermediate points leading to the next iterate:

(i) Fix $ x_b^{(k)}$ and update $ x_a^{(k)}.$ Set $y_b^{(k)}  \leftarrow \Phi_b x_b^{(k)}$ and $u_a^k  \leftarrow  y- y_b^{(k)}$ (so $ y= u_a^{(k)}+ y_b^{(k)}$).  Solve the system $ u_a^{(k)} = \Phi_a x_a $ to get $ x_a = \Phi_a^T u_a^{(k)}$. Performing thresholding yields $  \widetilde{x}^{(k)}_a={\cal H}_K(x_a)=  {\cal H}_K(\Phi_a^T u_a^{(k)})$  by which $x_a^{(k)} $ can be updated.

(ii) Fix $ \widetilde{x}_a^{(k)}$ and update  $ x_b^{(k)}.$  Set $\widetilde{y}_a^{(k)} \leftarrow \Phi_a \widetilde{x}_a^{(k)}$ and $u_b^{(k)} \leftarrow  y- \widetilde{y}_a^{(k)}$ (so $ y= u_b^{(k)} + \widetilde{y}_a^{(k)}$).   Solve the system $ u_b^{(k)} = \Phi_b x_b $ to get $ x_b = \Phi_b^T u_b^{(k)}$. Performing thresholding yields $ \widetilde{x}^{(k)}_b = {\cal H}_K(x_b) = {\cal H}_K(\Phi_b^T u_b^{k})$ by which $x_b^{(k)}$ can be updated

We now formally state the algorithm which is referred to as the \emph{Two-block Splitting Alternating Algorithm} (TSAA).

\textbf{TSAA:} Input vector $ y\in \mathbb{R}^m,$ sparsity level $K,$ matrix $A=[\Phi_a, \Phi_b] \in \mathbb{R}^{m\times 2m}$ with  $\Phi_a, \Phi_b\in \mathbb{R}^{m\times m}$ being orthogonal, and integer number $\tau$ such that $K \leq \tau \leq 2K. $

\begin{itemize}

\item[S1] (Initialize) Give any initial point $ x^{(0)} = (x^{(0)}_a, x^{(0)}_b)\in \mathbb{R}^{2m} $ and any initial vectors $ y_b^{(0)}$ in $\mathbb{R}^m. $ Set $ k:=0$ (initial count number of iteration).

\item[S2] At the current iterate $x^{(k)}$ together with $y_b^{(k)}$, set
$$  u^{(k)}_a = y- y^{(k)}_b,
 ~ \widetilde{x}_a^{(k)} = {\cal H}_K(\Phi_a^T   u^{(k)}_a), ~  u^{(k)}_b = y- \Phi_a \widetilde{x}_a^{(k)},
   ~ \widetilde{x}_b^{(k)} = {\cal H}_K(\Phi_b^T   u^{(k)}_b). $$

 \item[S3] Let $ \widetilde{x}^{(k)} = (\widetilde{x}_a^{(k)}, \widetilde{x}_b^{(k)})$ and $d^{(k)}=A^T(y-Ax^{(k)}).$  Set  $ \Lambda^{(k)} = {\cal L}_{\tau}(\widetilde{x}^{(k)}) \cup {\cal L}_{2K-\tau} (d^{(k)}) $ and
\begin{equation} \label{2S2A003}   \widehat{x}^{(k)}= \mathop{\arg\min}_{x \in \mathbb{R}^{2m}}\{\|y- A x\|_2:  ~ \textrm{supp}(x) \subseteq \Lambda^{(k)}\}. \end{equation}

\item[S4] Set $ S^{(k)} =  {\cal L}_K(\widehat{x}^{(k)} )$ and
\begin{equation} \label{SSAA004}   (x_a^{(k+1)}, x_b^{(k+1)})=x^{(k+1)}: = \mathop{\arg\min}_{x  \in \mathbb{R}^{2m}} \{\|y-A x \|_2:  ~ \textrm{supp}(x) \subseteq S^{(k)}\}  .
\end{equation}
Set $ y^{(k+1)}_b:  = \Phi_b x_b^{(k+1)}  $. Replace $ k+1$ by $ k$ and repeat S2-S4 until a stopping criterion is met.
\end{itemize}

\begin{Rem} \label{rrem001} (i) To more efficiently chase a sparse point, we incorporate two shrinkages and orthogonal projections into the procedure as shown in S3 and S4 above. The first shrinkage is relatively loose, aiming primarily to reduce the error $\|y-Ax\|_2,$ while the second ensures that the iterate is $K$-sparse. (ii) Various stopping criteria can be employed in TSAA. For example, the algorithm may terminate when $\|y-Ax^{(k)}\|_2 \leq \varepsilon,$ where $\varepsilon > 0$  is a given tolerance, or after reaching the prescribed maximum number of iterations. (iii) The computational cost of TSAA is low. Steps S2 and S3 only involve matrix-vector products, and (\ref{2S2A003}) and (\ref{SSAA004}) are small least-squares problems with at most $2K$ and $K$ variables, respectively.
\end{Rem}

\section{Analysis of TSAA} \label{AnalSAA}
In this section,    we prove that the sequence  $\{ x^{(k)}\}_{k\geq 1} ,  $ generated by TSAA, converges to the solution of (\ref{Sparse-Sol-LS}) under some assumption. Several useful lemmas are needed for the purpose. The first one below was taken from \cite{ZL23}.

\begin{Lem}\label{HHB}  \cite{ZL23} For any vector $z\in \mathbb{R}^n$ and any $K$-sparse vector $x\in \mathbb{R}^n$, one has
$ \|x- {\cal H}_K(z)\|_2\leq \omega \|(x-z)_{S\cup S^*}\|_2,
$
where $\omega := \frac{\sqrt{5}+1}{2},$   $  S= \textrm{supp}(x)$ and $S^*  =  \textrm{supp}({\cal H}_K(z)) .$
\end{Lem}

\begin{Lem} \label{Lem-Ml1l2}
Let $ M\in \mathbb{R}^{\ell_1\times \ell_2}$ be a matrix with absolute entries $|m_{ij}|\leq \alpha $ for some number $ \alpha>0.$  Then for  any  $ u\in \mathbb{R}^{\ell_2},$ one has
$ \|Mu\|_2 \leq \frac{\ell_1+\ell_2}{2} \alpha \|u\|_2. $
\end{Lem}

\emph{Proof. } Let $ m^{(i)} = (m_{i1}, \dots, m_{i\ell_2}), i=1, \dots, \ell_1$ denote the rows of $M. $ Then
\begin{align*} \|Mu\|_2^2
  =  \sum_{i=1}^{\ell_1} (m^{(i)} u)^2
  \leq   \sum_{i=1}^{\ell_1} (\sum_{j=1}^{\ell_2} |m_{ij}|^2) \|u\|_2^2
     \leq \ell_1\ell_2 \alpha ^2 \|u\|_2^2,
  \end{align*}
which implies that $ \|Mu\|_2 \leq  \sqrt{\ell_1\ell_2} \alpha  \|u\|_2 \leq \frac{\ell_1+\ell_2}{2} \alpha  \|u\|_2 .$   \  \   $ \Box $

We now establish the next two lemmas that may be of independent interest.

\begin{Lem} \label{Lem-Ml1l3} Let $A=[\Phi_a, \Phi_b],$ where $ \Phi_a, \Phi_b \in \mathbb{R}^{m\times m}$ are orthogonal matrices. Let $ \Lambda \subseteq \{1, \dots, 2m\}$ be an index set with $|\Lambda|< m.$ Then for any $ u\in \mathbb{R}^{2m}$,     one has
$$ \left\|[(A^TA-I)u]_\Lambda\right\|_2 \leq \mu(A)\left(|\Lambda| \cdot \|u_\Lambda\|_2+ m \|u_{\overline{\Lambda}}\|_2\right),
 $$
 where  $ \overline{\Lambda} =  \{1, \dots, 2m\} \setminus \Lambda. $
\end{Lem}

\emph{Proof.}    Denoted by
$ M: = A^TA-I = \left[
                 \begin{array}{cc}
                   0 & \Phi_a^T \Phi_b \\
                   \Phi_b^T \Phi_a & 0 \\
                 \end{array}
               \right]. $
               Let $ \Lambda $ be the index set given in the lemma. Still, let $ m^{(i)}$ denote the $i$th row of $M$ and
 $ I^{(i)} = \textrm{supp} (m^{(i)}).$  Note that every nonzero entry of $M$  is bounded by $ \mu(A), $ and that every row of $M$ contains at most $m$ nonzeros and hence $ |I^{(i)}| \leq m. $  For any $ u\in \mathbb{R}^{2m},$ we have
\begin{align}  \label{mmuu} |m^{(i)}u| &= | (m^{(i)})_\Lambda u_\Lambda + (m^{(i)})_{\overline{\Lambda}} u_{\overline{\Lambda}}|
   = |(m^{(i)})_{\Lambda\cap I^{(i)}} u_{\Lambda\cap I^{(i)}} + (m^{(i)})_{\overline{\Lambda}\cap I^{(i)}} u_{\overline{\Lambda}\cap I^{(i)}}|\nonumber \\
  & \leq \|(m^{(i)})_{\Lambda\cap I^{(i)}}\|_2\| u_{\Lambda\cap I^{(i)}}\|_2 + \|(m^{(i)})_{\overline{\Lambda}\cap I^{(i)}}\|_2 \| u_{\overline{\Lambda}\cap I^{(i)}}\|_2.
\end{align}
Notice that  for any vector $ v $ with $\ell $ nonzero entries which are bounded as $|v_i| \leq \beta $, one has  $ \|v\|_2 \leq \sqrt {\ell } \beta.$ Thus
$\|(m^{(i)})_{\Lambda\cap I^{(i)}}\|_2 \leq \sqrt{|\Lambda\cap I^{(i)}|}\mu(A)  $ and $ \|(m^{(i)})_{\overline{\Lambda}\cap I^{(i)}}\|_2 \leq \sqrt{|\overline{\Lambda}\cap I^{(i)}|}\mu(A). $ We also note that $ \| u_{\Lambda\cap I^{(i)}}\|_2 \leq  \| u_{\Lambda}\|_2$ and  $  \| u_{\overline{\Lambda}\cap I^{(i)}}\|_2 \leq \|u_{\overline{\Lambda}}\|_2. $ Therefore, it follows from (\ref{mmuu}) that
\begin{equation} \label{Case-a}  |m^{(i)}u|\leq \mu(A) \left(\sqrt{|\Lambda\cap I^{(i)}|}\| u_{\Lambda}\|_2 +  \sqrt{|\overline{\Lambda}\cap I^{(i)}|} \|u_{\overline{\Lambda}}\|_2 \right).
\end{equation}
There is an index $i\in \Lambda$ such that  $ |m^{(i)} u|= \max_{j\in \Lambda} |m^{(j)}u|.$
By using (\ref {Case-a}), we obtain
\begin{align*}  \label{matrix-bound-02}\|(Mu)_\Lambda\|_2    \leq  \sqrt{|\Lambda|} | |m^{(i)} u|  & \leq \mu(A)  \sqrt{|\Lambda |} \left(\sqrt{|\Lambda\cap I^{(i)}|}\| u_{\Lambda}\|_2 +  \sqrt{|\overline{\Lambda}\cap I^{(i)}|}\|u_{\overline{\Lambda}}\|_2 \right)\\
& \leq  \mu(A) \left(|\Lambda|| u_{\Lambda}\|_2+ m \|u_{\overline{\Lambda}}\|_2\right) ,\end{align*}
which follows from    $ |\Lambda\cap I^{(i)}| \leq  |\Lambda| $ and
$ \sqrt{|\Lambda |} \sqrt{|\overline{\Lambda}\cap I^{(i)}|} \leq   (|\Lambda| +|I^{(i)}|)/2\leq m. $   ~~ $ \Box $

\begin{Lem} \label{Lem-M1114} Let $A=[\Phi_a, \Phi_b],$ where $ \Phi_a, \Phi_b \in \mathbb{R}^{m\times m}$ are two orthogonal matrices. Let $x ^*$ be the $K$-sparse solution to the system $ y=Ax$ and $K< 1/\mu(A).$ Let $ \Lambda \subseteq \{1, \dots, 2m\}$ be an index set with cardinality $K\leq |\Lambda| < 1/\mu(A). $
Let \begin{equation}\label{proj-orth} z^+ = \arg \min_{z\in \mathbb{R}^{2m}}\{\| y-Az\|_2: ~ \textrm{supp} (z) \subseteq \Lambda\}.
\end{equation}
 Then
 \begin{equation}\label{Error-Fin}  \| z^+ - x^*\|_2 \leq \sqrt{1+ \left(\frac{m\mu(A)}{1-|\Lambda|\mu(A)}\right)^2   } \|( z^+ - x^*)_{\overline{\Lambda}}\|_2.
 \end{equation}
\end{Lem}
where $ \overline{\Lambda}= \{1, \dots, 2m\} \setminus \Lambda.$

\emph{Proof.} As $z^+$ is the solution to (\ref{proj-orth}), at  this point the components of the gradient of $ \|y-Az\|_2^2$ indexed by $ \Lambda$ vanish. Namely, $[A^T (y-Az^+)]_\Lambda =0. $ Substituting $ y=Ax^*$ into this equality leads to
$[A^T A(x^*-z^+)]_\Lambda =0, $
and hence
$ -(z^+ - x^*)_\Lambda = [(A^TA-I) (z^+ - x^*)]_\Lambda . $
 Thus, by Lemma \ref{Lem-Ml1l3}, one has
\begin{align*} \|(z^+ - x^*)_\Lambda\|   & = \|[(A^TA-I) (z^+ - x^*)]_\Lambda \|_2 \\
& \leq \mu(A) \left( |\Lambda|\|(z^+ - x^*)_\Lambda\|_2 + m \|(z^+ - x^*)_{\overline{\Lambda}}\|_2\right),
\end{align*}
which together with $\|z^+ - x^*\|_2^2 = \|(z^+ - x^*)_\Lambda\|_2^2 + \|(z^+ - x^*)_{\overline{\Lambda}}\|_2^2 $ implies (\ref{Error-Fin}).  ~~ $ \Box$

The lemma below provides a condition for the uniqueness of solution to (\ref{Sparse-Sol-LS}).

\begin{Lem}\label{Lem-M-classic} \cite{E10} If $ x^*$ is a $K$-sparse solution to the linear system $ Ax=y$ and $ K< \frac{1}{2}(1+\frac{1}{\mu(A)}),$ then $ x^*$ must be the unique sparsest solution to this system.
\end{Lem}

 We are ready to show that the sequence $\{x^{(k)}\}_{k\geq 1},$ generated by TSAA, converges to the sparsest solution of the linear system under a  condition expressed in $\mu(A).$

\begin{Thm} \label{Thm-Mu-Co} Let $ A =[\Phi_a, \Phi_b],$ where $ \Phi_a, \Phi_b\in \mathbb{R}^{m\times m}$ are orthogonal matrices. Suppose that $ x^*= (x^*_a, x_b^*)$ is the $K$-sparse solution to the system $ y= Ax$ and that
\begin{equation} \label{Mu-Condition} K < \frac{1}{2\mu (A)} \min\left\{c,  \frac{c(1-c)^2}{m^2} \left( \frac{1}{\mu (A)}\right)^2\right\} \end{equation}
where $ c= \frac{\sqrt{2}}{3\omega^3}<1 $ ($\omega = \frac{\sqrt{5}+1}{2}$). Let $ x^{(0)}  \in \mathbb{R}^{2m} $ and $ y^{(0)}_b \in \mathbb{R}^m $ be any given initial vectors. Then the sequence $ \{ x^{(k)}=(x^{(k)}_a, x^{(k)}_b)\}_{k\geq 1}, $   generated by TSAA, satisfies
  $$ \|x^{(k+1)} - x^*\|_2 \leq  \rho \|x_b^{(k)}-x^*_b \|_2 $$  for all $ k\geq 1,$
where {\small \begin{equation}\label{rrhhoo} \rho = \frac{3K\omega^3 \mu(A) }{2}  \sqrt{\left[1+ \left(\frac{m\mu(A)}{1-K\mu(A)}\right)^2\right] \left[1+ \left(\frac{m\mu(A)}{1-2K\mu(A)}\right)^2\right]   \left[1+ \left(  \frac{3K\omega \mu(A)}{2} \right)^2\right]}, \end{equation}}
which is strictly smaller than 1 under (\ref{Mu-Condition}). Thus the sequence $\{x^{(k)}\}_{k\geq 1}  $  converges to  $x^*,$ the unique sparsest solution of the linear system.
\end{Thm}

\emph{Proof.}  Condition (\ref{Mu-Condition}) implies $ K< \frac{1}{2}(1+\frac{1}{\mu(A)})$ (see Remark \ref{Rem-Sect-01} for details) which, by Lemma \ref{Lem-M-classic}, ensures that $ x^* = (x^*_a, x^*_b) $ is the unique sparsest solution to the system $y=Ax.$ The vector $y$ can be written as  $ y=y^*_a + y^*_b, $ where $ y^*_a = \Phi_a x^*_a $ and $ y^*_b= \Phi_b x^*_b. $ Given $x^{(k)}= (x_a^{(k)}, x_b^{(k)}),$ TSAA generates  $x^{(k+1)}= (x_a^{(k+1)}, x_b^{(k+1)}) $ by performing its steps S2-S4.
Let $x^{(k)}_a,  x^{(k)}_b,  \widetilde{x}^{(k)}_a, \widetilde{x}^{(k)}_b, u^{(k)}_a,  u^{(k)}_b  $  be defined as in TSAA.  Denote by  $ z_a^{(k)} =\Phi^T_a u_a^{(k)} $ and $  z_b^{(k)} = \Phi^T_b u_b^{(k)}. $
We see from S2 that for all $ k\geq 1,$   $ u_b^{(k)}+\Phi_a \widetilde{x}_a^{(k)} =y = y^*_a + y^*_b = \Phi_a x_a^* + \Phi_b x_b^*.$ Thus
\begin{align}
 z_b^{(k)}-x_b^* &  = \Phi^T_b u_b^{(k)}-x_b^*  = \Phi^T_b (\Phi_a x_a^* + \Phi_b x_b^* - \Phi_a \widetilde{x}^{(k)}_a) - x_b^*  \nonumber \\
 & = \Phi^T_b\Phi_a (x^*_a-\widetilde{x}_a^{(k)}),
    \label{zbxb}
 \end{align}
 where the last equality follows from the orthogonality of $ \Phi_b.$ From S2 of TSAA, we see that for all $ k\geq 1,$ $u_a^{(k)}+y_b^{(k)} =y = \Phi_a x_a^* + \Phi_b x_b^*    $ and $ y_b^{(k)}= \Phi_b x_b^{(k)}. $  Thus
 \begin{align}
 z_a^{(k)}-x_a^* & = \Phi^T_a u_a^{(k)}-x_a^*  = \Phi^T_a (\Phi_a x^*_a +\Phi_b x^*_b-\Phi_bx_b^{(k)}) -x_a^* \nonumber \\
    & = \Phi^T_a\Phi_b (x^*_b-x_b^{(k)}),   \label{zaxa}
 \end{align}
 where the last equality follows from the orthogonality of $ \Phi_a.$ Denote by $  S_a= \textrm{supp}(x^*_a)$,  $ S_b = \textrm{supp}(x^*_b),$  $\Omega^{(k)}_a= \textrm{supp} (\widetilde{x}_a^{(k)})= \textrm{supp} ({\cal H}_K(z_a^{(k)}) ) $ and $ \Omega_b^{(k)} = \textrm{supp} (\widetilde{x}_b ^{(k)})= \textrm{supp}({\cal H}_K (z_b^{(k)})).$  Thus  $|\Omega^{(k)}_a|,   |\Omega_b^{(k)}|\leq K.$ Denote by $\Gamma^{(k)}_b =\textrm{supp} (x^{(k)}_b).$ Then, by
  (\ref{zaxa}) and Lemma \ref{HHB}, one has
 \begin{align} \label{OERR-A}
\|\widetilde{x}^{(k)}_a- x_a^*\|_2  & = \|{\cal H}_K(z_a^{(k)})-x_a^*\|_2 \nonumber\\
  & \leq  \omega \| (z_a^{(k)}-x_a^* )_{ {\small S_a\cup \Omega_a^{(k)}} } \|_2 =  \omega \|  [\Phi^T_a\Phi_b (x^*_b-x_b^{(k)}) ]_{S_a\cup \Omega_a^{(k)}} \|_2  \nonumber\\
   & = \omega \|  [(\Phi_a)_{S_a\cup\Omega_a^{(k)}} ]^T  \Phi_b (x^*_b-x_b^{(k)}) \|_2 \nonumber\\
& = \omega \|  [(\Phi_a)_{ S_a\cup\Omega_a^{(k)}} ]^T (\Phi_b)_ {S_b\cup\Gamma_b^{(k)}} (x^*_b-x_b^{(k)})_{S_b\cup\Gamma_b^{(k)}} \|_2,
\end{align}
where the last equality follows from the  fact $ \textrm{supp}(x^*_b-x_b^{(k)})\subseteq \textrm{supp}(x^*_b) \cup \textrm{supp}(x_b^{(k)})= S_b \cup \Gamma_b^{(k)}.$
Note that every entry of the matrix $ M= \left[(\Phi_a)_{S_a\cup \Omega_a^{(k)}}\right]^T (\Phi_b)_ {S_b\cup\Gamma^{(k)}_b}$ is the inner product of  a column of $ \Phi_a$ and a column of $ \Phi_b$. Thus the entries of this matrix are bounded by $ \mu(A).$
Thus, by Lemma \ref{Lem-Ml1l2}, we immediately have that
\begin{align} \label{mut-Bou-A}    \| [(\Phi_a)_{ S_a\cup\Omega_a^{(k)}}]^T   &   (\Phi_b)_ {S_b\cup\Gamma_b^{k}} (x^*_b-x_b^{(k)})_{S_a\cup\Gamma_b^{(k)}} \|_2   \nonumber \\
 & \leq \frac{1}{2} (|S_a\cup\Omega_a^{(k)}|+|S_b\cup\Gamma_b^{(k)}|) \mu(A) \|(x^*_b-x_b^{(k)})_ { S_b\cup\Gamma_b^{(k)}}\|_2  \nonumber \\
   & \leq \left(3K/2\right)\mu(A) \|x^*_b-x_b^{(k)}\|_2,
 \end{align}
 where the last inequality follows from the fact
 $  |S_a\cup\Omega^{(k)}_a|+|S_b\cup\Gamma^{(k)}_b| \leq |\Omega^{(k)}_a|+ |\Gamma^{(k)}_b|+|S_a|+|S_b| \leq 3K. $
Merging (\ref{OERR-A}) and (\ref{mut-Bou-A}) yields
 \begin{align} \label{Err-BOU1} \|\widetilde{x}^{(k)}_a- x_a^*\|_2 \leq \omega \mu(A) \left(3K/2\right)\|x^*_b-x_b^{(k)}\|_2.
 \end{align}
Similarly,  by (\ref{zbxb}) and   Lemmas \ref{HHB}, one has
\begin{align*}
\|\widetilde{x}^{(k)}_b- x_b^*\|_2 & = \|{\cal H}_K(z_b^{(k)})-x_b^*\|_2 \nonumber  \\
&\leq \omega \|(z_b^{(k)}-x_b^*)_{ S_b\cup\Omega^{(k)}_b}\|_2 \nonumber \\
& = \omega \| [\Phi^T_b\Phi_a (x^*_a-\widetilde{x}_a^{(k)})]_{ S_b\cup\Omega^{(k)}_b}
\|_2  \nonumber \\
& = \omega \|  [(\Phi_b)_{S_b\cup\Omega^{(k)}_b}]^T (\Phi_a)_{ S_a\cup\Omega^{(k)}_a} (x^*_a-\widetilde{x}_a^{(k)})_{ S_a\cup\Omega^{(k)}_a} \|_2.
\end{align*}
By Lemmas  \ref{Lem-Ml1l2}, we have that
\begin{align} \label{OERR-B}
\|\widetilde{x}^{(k)}_b- x_b^*\|_2
& \leq \frac{1}{2}\omega \mu(A)\left(| S_b\cup\Omega^{(k)}_b|+| S_a\cup\Omega^{k}_a|\right)\|(x^*_a-\widetilde{x}_a^{(k)})_{ S_a\cup\Omega^{(k)}_a} \|_2   \nonumber \\
& \leq \omega \mu(A)\left(3K/2\right)\|x^*_a-\widetilde{x}_a^{(k)}\|_2,
\end{align}
where the last inequality follows from
$ |S_b\cup\Omega^{(k)}_b|+|S_a\cup\Omega^{(k)}_a| \leq  |\Omega^{(k)}_b|+|\Omega^{(k)}_a |+ | S_a| + |S_b|\leq 3K. $
Merging (\ref{Err-BOU1}) and (\ref{OERR-B})  yields
\begin{equation} \label{MERG1314} \|\widetilde{x}^{(k)}_b- x_b^*\|_2 \leq \left[\omega \mu(A) \left(3K/2\right)\right]^2 \|x^*_b-x_b^{(k)}\|_2 .
\end{equation}
Denote by $ u= {\cal H}_{\tau} (\widetilde{x}^{(k)}),$ where $\widetilde{x}^{(k)} = (\widetilde{x}_a^{(k)},\widetilde{x}_b^{(k)} )$ and  $\tau$ is the integer number given in TSAA. By Lemma \ref{HHB}, we have $ \|u-x^*\|_2 \leq \omega \|(\widetilde{x}^{(k)}-x^*)_{S\cup \Lambda}\|_2 \leq \omega \|\widetilde{x}^{(k)}-x^*\|_2, $ where
$ S=\textrm{supp}(x^*)$ and $ \Lambda = \textrm{supp}(u).$ Combing with (\ref{Err-BOU1}) and  (\ref{MERG1314}) yields
\begin{equation}\label{1A2702-a}  \|u-x^*\|_2 \leq \omega^2 \mu(A)\left( 3K/2\right)  \sqrt{1+ \left[3K\omega \mu(A)/2\right]^2}   \|x_b^{(k)}-x^*_b\|_2. \end{equation}
By the structure of TSAA, the $2K$-sparse vector $ \widehat{x}^{(k)}$ is the solution to the least-squares problem (\ref{2S2A003}).  Note  that $ \textrm{supp}(u) \subseteq {\cal L}_\tau (\widetilde{x}^{(k)}) \subseteq \Lambda^{(k)}=  {\cal L}_\tau (\widetilde{x}^{(k)})  \cup {\cal L}_{2K-\tau}(d^{(k)})  , $ where $ d^{(k)}= A^T (y-A x^{(k)}).$ Clearly, $|\Lambda^{(k)}| \leq 2K < \frac{1}{\mu(A)},$ where the second inequality follows from (\ref{Mu-Condition}). Thus it follows from Lemma \ref{Lem-M1114} that
 $$ \|\widehat{x}^{(k)}-x^*\|_2 \leq \rho_1 \|(\widehat{x}^{(k)}-x^*)_{\overline{\Lambda^{(k)}}} \|_2 = \rho_1 \|(u-x^*)_{\overline{\Lambda^{(k)}}}\|_2\leq \rho_1 \|u-x^* \|_2,
$$
  where $\rho_1 = \sqrt{1+ \left(\frac{m\mu(A)}{1-2K\mu(A)}\right)^2  },$ and the equality follows from   $(\widehat{x}^{(k)})_{\overline{\Lambda^{(k)}}} =0 = (u)_{\overline{\Lambda^{(k)}}}. $

Denote by $ v= {\cal H}_K (\widehat{x}^{(k)}). $ From Lemma \ref{HHB} and the  inequality above, we have
 \begin{equation} \label{ER-ZYB-25} \|v-x^*\|_2\leq \omega \|(\widehat{x}^{(k)}-x^*)_{S\cup S^{(k)}}\|_2\leq \omega\rho_1 \|u-x^* \|_2, \end{equation}
 where $S=\textrm{supp} (x^*)  $ and $ S^{(k)} =\textrm{supp} (v)$ with $ |S^{(k)}|\leq K.$  As $ x^{(k+1)} = (x_a^{(k+1)}, x_b^{(k+1)})$ is the solution to the problem (\ref{SSAA004}), by Lemma \ref{Lem-M1114}, we deduce that
$$ \|x^{(k+1)} - x^* \|_2 \leq \rho_2  \|(x^{(k+1)} - x^*)_{\overline{S^{(k)}}}\|_2=  \rho_2 \|(v - x^*)_{\overline{S^{(k)}}}\|_2 \leq \omega \rho_1\rho_2  \|u-x^* \|_2, $$
where  $\rho_2= \sqrt{1+ \left(\frac{m\mu(A)}{1-K\mu(A)}\right)^2 } $,   the equality is due to $ (x^{(k+1)})_{\overline{S^{(k)}}} =0 =(v)_{\overline{S^{(k)}}}$, and the last inequality follows from (\ref{ER-ZYB-25}).
Merging this inequality with (\ref{1A2702-a}) yields
\begin{equation}\label{ERB2024} \|x^{(k+1)} - x^*\|_2 \leq  \rho \|x_b^{(k)}-x^*_b\|_2  \end{equation}
where $$\rho: = \omega^3 \rho_1\rho_2 \mu(A)\left( 3K/2\right)  \sqrt{1+ \left[3K\omega \mu(A)/2\right]^2}, $$
which is exactly the constant in (\ref{rrhhoo}). It is not difficult to verify that  $ \rho <1$ under   (\ref{Mu-Condition}). In fact,   (\ref{Mu-Condition}) implies that $ 1-2K\mu(A) >0$ and $\frac{3K\omega \mu(A)}{2} < 1 $  which, together with  $   \frac{1}{1-K\mu(A)} <  \frac{1}{1-2K\mu(A)},  $
 implies that
 $$  \rho_1\rho_2 \sqrt{1+ \left( \frac{3K\omega \mu(A)}{2}\right)^2}  < \sqrt{2} \left[1+ \left(\frac{m\mu(A)}{1-2K\mu(A)}\right)^2\right] .$$
 Thus
 $$ \rho <  \frac{3\sqrt{2}K\omega^3 \mu(A)}{2}  \left[1+ \left(\frac{m\mu(A)}{1-2K\mu(A)}\right)^2\right] . $$
 If $ \frac{m\mu(A)}{1-2K\mu(A)} \leq 1,$  then $  \rho \leq 3\sqrt{2}K\omega^3 \mu(A) <1 $ under (\ref{Mu-Condition}); otherwise, we have
 $$ \rho \leq  3\sqrt{2}K\omega^3 \mu(A) \left(\frac{m\mu(A)}{1-2K\mu(A)}\right)^2 \leq \frac{3\sqrt{2}K\omega^3 \mu(A)^3 m^2} {(1-c)^2} <1, $$
 where the second inequality follows from  $ 1-2K\mu(A) > 1-c$ due to  (\ref{Mu-Condition}) which implies $ 2K\mu(A) <  c <1, $
 and the last inequality is implied from (\ref{Mu-Condition}). Since $ \|x^*_b-x_b^{(k)}\|_2  \leq \|x^*-x^{(k)}\|_2 ,$  it follows from (\ref{ERB2024}) that the sequence $\{x^{(k)}\}$ converges to the sparse solution $ x^*$  of the linear system.  \  \  $\Box$

\begin{remark} \label{Rem-Sect-01} The above theorem shows that
(\ref{Mu-Condition}) is a sufficient condition for the global convergence of TSAA.  Note that $$ \frac{1}{2\mu (A)} \min\left\{c,  \frac{c(1-c)^2}{m^2} \left( \frac{1}{\mu (A)}\right)^2\right\} \leq \frac{c}{2\mu (A)} < \frac{1}{2\mu (A)} < \frac{1}{2}\left(1+\frac{1}{\mu (A)}\right),$$ where $c$ is the constant given in Theorem \ref{Thm-Mu-Co}. Thus (\ref{Mu-Condition}) is more conservative than the one in Lemma \ref{Lem-M-classic}. Given a matrix $A$, the right-hand side of (\ref{Mu-Condition}) is easy to compute. Clearly, the smaller the value of  $\mu(A)$, the larger the quantity of the right-hand side of (\ref{Mu-Condition}), and thus the broader the class of linear systems that satisfy  (\ref{Mu-Condition}).
\end{remark}

\section{Multi-block splitting alternating algorithm}
In this section, we consider the  multi-block case where the matrix is formed by concatenating more than two orthogonal matrices.
 For this case, the SAA is referred to as the \emph{Multi-block Splitting Alternating Algorithm} (MSAA) which is formally described as follows.

\textbf{MSAA} Input $A=[\Phi_1, \dots, \Phi_p] \in \mathbb{R}^{m\times pm},$ where $ p>2$ and $ \Phi_i \in \mathbb{R}^{m\times m}, i=1, \dots, p$ are orthogonal matrices. Input  $ y\in \mathbb{R}^m,$ sparsity level $K, $ and integer number $\tau$ such that $K \leq \tau \leq 2K. $

\begin{itemize}

\item[S1] (Initialize) Give any initial point $ x^{(0)} = (x^{(0)}_1, \dots, x^{(0)}_p)\in \mathbb{R}^{pm} $  and any initial vectors $ y_i^{(0)}\in   \mathbb{R}^m $ for $ i=2, \dots, p.$ Set $ k:=0.$

\item[S2] Given vectors $x^{(k)} $ and $  y_i^{(k)} \in \mathbb{R}^m$ for $ i=2, \dots, p,$  set $u_1^{(k)} = y-\sum_{j=2}^{p} y_j^{(k)}$  and perform the following loop:

 \begin{itemize}
   \item[]for $ i=1, \dots, p-1, $ do
    \item[] $ \widetilde{x}_i^{(k)}={\cal H}_K(\Phi_i^T u_i^{(k)});  ~ \widetilde{y}_i^{(k)} = \Phi_i \widetilde{x}_i ^{(k)}; ~  u_{i+1}^{(k)}= y-\sum_{j=1}^{i} \widetilde{y}_j^{(k)} - \sum_{j=i+2}^p y_j^{(k)} $
\item[] end
\end{itemize}
 Set $ \widetilde{x}_p^{(k)}={\cal H}_K(\Phi_p^T u_p^{(k)}). $

\item[S3] Let $ \widetilde{x}^{(k)} = (\widetilde{x}_1^{(k)},\dots, \widetilde{x}_p^{(k)})$ and $d^{(k)}=A^T(y-Ax^{(k)}).$  Set  $ \Lambda^{(k)} = {\cal L}_{\tau}(\widetilde{x}^{(k)}) \cup {\cal L}_{2K-\tau} (d^{(k)}) $
 and
\begin{equation} \label{MSAA03}   \widehat{x}^{(k)}= \mathop{\arg\min}_{x \in \mathbb{R}^{pm}}\{\|y- A x\|_2: ~ \textrm{supp}(x) \subseteq \Lambda^{(k)}\}. \end{equation}

\item[S4] Set $ S^{(k)} =  {\cal L}_K(\widehat{x}^{(k)} )$ and
\begin{equation} \label{MSAA04}   (x_1^{(k+1)}, \dots, x_p^{(k+1)}) = x^{(k+1)}:= \mathop{\arg\min}_{x  \in \mathbb{R}^{pm}} \{\|y-A x \|_2:  ~ \textrm{supp}(x) \subseteq S^{(k)}\}  . \end{equation}
Set $ y_i^{(k+1)} = \Phi_i x_i^{(k+1)},  i= 2, \dots, p. $
Replace $ k+1$ by $ k,$ and repeat S2-S4 until a stopping criterion is met.\\
\end{itemize}

The comment similar to Remark \ref{rrem001} is valid for MSAA. The loop at S2 iteratively and alternately generates the intermediate point $ \widetilde{x}^{(k)} =  (\widetilde{x}_1^{(k)}, \dots, \widetilde{x}_p^{(k)})
$ from the current one
$x^{(k)}. $
Starting with $  u_1^{(k)} : = y-\sum_{j=2}^p y_j^{(k)}$,   the algorithm  generates
$  \widetilde{x}_1^{(k)}, $ $\widetilde{y}_1^{(k)},$ and $ u_2^{(k)}, $
then
$  \widetilde{x}_2^{(k)},  \widetilde{y}_2^{(k)} $ and  $ u_3^{(k)}, $ and continues
until  $ u^{(k)}_p$ and $ \widetilde{x}_p^{(k)}$ are generated at the end of loop.  The vectors
$$ u^{(k)}_i = y-
\sum_{j<i} \widetilde{y}_j ^{(k)} -
\sum_{j> i} y_i^{(k)}, ~  i=1, \dots, p$$ can be seen as the approximation to the optimal partition $ y^*_1, \dots, y^*_p $
  of $y$, which can be written as
$ y^*_i = y-
\sum_{j<i} y^*_j - \sum_{j> i} y^*_j, $ $ i=1, \dots, p.$ It is also interesting to note that when $ \tau = 2K$,  the algorithm does not  use the gradient information $ d^{(k)} = A^T(y-Ax^{(k)})$ of the metric $\|y-Ax\|_2^2 $ at $ x^{(k)}.$  In this case, $ \Lambda^{(k)}= {\cal L}_{2K} (\widetilde{x}^{(k)}). $

\section{Analysis of MSAA}

In this section, we show that under some condition, the sequence generated by MSAA converges to the sparsest solution of the linear system.   Before stating the main result for MSAA, let us introduce the following norm of the vector $z = (z_1, \dots, z_p)\in \mathbb{R}^{pm},$ where each $ z_i \in \mathbb{R}^m:$
$$  \|z\|_{(p,\infty)} = \max_{1\leq i\leq p} \|z_i\|_2, $$
 which is helpful in the analysis of MSAA. We also need the following generalized versions of Lemmas \ref{Lem-Ml1l3} and \ref{Lem-M1114}, respectively.

 \begin{Lem} \label{Lem-MSAA01} Let $A=[\Phi_1, \Phi_2, \dots, \Phi_p],$ where $ \Phi_i \in \mathbb{R}^{m\times m}, i=1, \dots, p $  are orthogonal matrices. Let $ \Lambda \subseteq \{1, \dots, pm\}$ be an index set with $|\Lambda|< m.$ Then for any vector $ u\in \mathbb{R}^{pm}$, one has
\begin{equation} \label {DDMSAA} \left\|[(A^TA-I)u]_\Lambda\right\|_2 \leq \mu(A)(|\Lambda| \cdot \|u_\Lambda\|_2+ \frac{1}{2}pm \|u_{\overline{\Lambda}}\|_2), \end{equation}
where $ \overline{\Lambda}= \{1, \dots, pm\}\setminus \Lambda.$
\end{Lem}

Note that every row of  $ A^TA-I$ contains at most $(p-1)m $ nonzero entries.  Using a proof similar to that of  Lemma \ref{Lem-Ml1l3} will yield  (\ref{DDMSAA}). The proof is omitted here.

\begin{Lem} \label{Lem-MSAA02} Let $A=[\Phi_1, \dots, \Phi_p]\in \mathbb{R}^{m\times pm},$ where $ \Phi_i \in \mathbb{R}^{m\times m}, i=1, \dots, p $  are orthogonal matrices. Let $x ^*$ be the $K$-sparse solution to the system $ y=Ax$ and $K< 1/\mu(A).$ Let $ \Lambda \subseteq \{1, \dots, pm\}$ be an index set with cardinality $ |\Lambda|$ satisfying $K\leq |\Lambda| < 1/\mu(A). $
Let $z^+ =\mathop{\arg\min}_{z\in \mathbb{R}^{pm}}\{\| y-Az\|_2: \textrm{supp} (z) \subseteq \Lambda\}.
$
 Then
 \begin{equation}\label{Error-F}  \|z^+-x^*\|_2 \leq \sqrt{1+ \left(\frac{pm\mu(A)}{2(1-|\Lambda|\mu(A))}\right)^2   } \|(z^+-x^*)_{\overline{\Lambda}}\|_2,
 \end{equation}
 where $ \overline{\Lambda} = \{1, \dots, pm\}\setminus \Lambda. $
\end{Lem}

Using Lemma \ref{Lem-MSAA01} and the  similar proof of Lemma \ref{Lem-M1114},  one can obtain the error bound (\ref{Error-F}) immediately. The details are omitted.

\begin{Thm} \label{Main-Thm-Mu-Co}  Let  $A=[\Phi_1, \dots, \Phi_p] \in \mathbb{R}^{m\times pm}$, where $ \Phi_i \in \mathbb{R}^{m\times m},$ $i=1, \dots, p$ are orthogonal matrices.  Suppose that the linear system $y= Ax $ has a $K$-sparse solution $ x^* \in \mathbb{R}^{pm}  $ and \begin{equation} \label{Multi-Cond}   K < \frac{1}{\mu(A) } \min\left\{\tau_1, \frac{ \tau_2}{m^2}\left(\frac{1}{\mu(A)}\right)^2\right\},
\end{equation} where $ \tau_1 = \frac{2}{\omega(1+2\omega^2(3p-2)\sqrt{p})}   $ ($< 1/5$) and $\tau_2 =  \frac{2(2- \omega \tau_1)(1-3\tau_1)^2}{\omega^3 (3p-2)p^{5/2}} $ are constants with $ \omega = (\sqrt{5}+1)/2.$
 Let $ x^{(0)}  \in \mathbb{R}^{pm} $ and $ y^{(0)}_i \in \mathbb{R}^m $ $(i=2, \dots , p)$ be any given initial vectors. Then sequence    $  \{ x^{(k)}\}_{k\geq 1} $ generated by MSAA satisfies that
  $$    \|x^{(k+1)}-x^* \|_2 \leq  \eta  \|x^{(k)} -x^*\|_{(p,\infty)},  $$
 where $ \eta $ is a positive constant given as
 $$   \eta =  \frac{\omega^3 K \mu(A)  (3p-2)\sqrt{p}} { 2- \omega K \mu(A)} \sqrt{ \left[1+ \left(\frac{pm\mu(A)}{2(1-3K\mu(A))}\right)^2\right] \left[1+ \left(\frac{pm\mu(A)}{2(1-K\mu(A))}\right)^2\right] }  <1.$$
Thus the sequence $\{x^{(k)}\}_{k\geq 1}  $  converges to  $x^* $ which, under (\ref{Multi-Cond}), is the unique sparsest solution to the  linear system.
\end{Thm}

\emph{Proof.} Suppose that the sequence
$\{x^{(k)}= (x^{(k)}_1, \dots, x^{(k)}_p)\}_{k\geq 1}, $ with $ x_i^{(k)} \in \mathbb{R}^m, $ is generated by MSAA. Let $\{\widetilde{x}^{(k)}, \widetilde{y}^{(k)}, \widehat{x}^{(k)}, x^{(k+1)} \}$ be given as in MSAA. We now prove that the sequence $ \{x^{(k)}\}_{k\geq 1} $ converges to $ x^*.$
For notational convenience, we define
$w_i^{(k)} = \Phi_i^T u_i^{(k)},  $ $ i=1, \dots, p,$  where $ u_i^{(k)}$'s are the vectors generated at S2 of the algorithm.
 We first provide an upper bound for  $ \|\widetilde{x}^{(k)}- x^*\|_{(p,\infty)} . $    Note that
$$ \|\widetilde{x}^{(k)}- x^*\|_{(p,\infty)} = \max_{1\leq i \leq p}\|\widetilde{x}_i^{(k)}- x_i^*\|_2 = \|\widetilde{x}_q^{(k)}- x_q^*\|_2 $$ for some $q\in \{1, \dots, p\}.$
By the structure of MSAA, for all $ k\geq 1,$ one has $\widetilde{y}_j^{(k)} =\Phi_j \widetilde{x}_j^{(k)}$ for $ j=1, \dots, p-1$ and $y_j^{(k)}=  \Phi_j x_j^{(k)}$ for $ j=2, \dots, p.$
By definition, $ u_i^{(k)} = y-\sum_{j< i}\widetilde{y}_j^{(k)} -\sum_{j>i}y_j^{(k)} $
which, together with  $ y= \Phi_qx^*_q+  \sum_{j<q} \Phi_j x^*_j + \sum_{j>q} \Phi_j x^*_j, $  implies that
\begin{align}
w_q^{(k)}- x_q^*  & =  \Phi_q^T u_q^{(k)} -x_q^*  = \Phi_q^T \Big{(} y - \sum_{j < q} \widetilde{y}_j^{(k)} - \sum_{j > q} y_j^{(k)}\Big{)}  -x_q^* \nonumber \\
 & = \Phi_q^T \Big{(}\Phi_qx^*_q+  \sum_{j<q} \Phi_j x^*_j + \sum_{j>q} \Phi_j x^*_j - \sum_{j < q} \Phi_j \widetilde{x}_j^{(k)} - \sum_{j > q} \Phi_j x_j^{(k)} \Big{)}  -x_q^* \nonumber \\
& =  \sum_{j< q} \Phi_q^T \Phi_j (x^*_j - \widetilde{x}_j^{(k)})  + \sum_{j > q} \Phi_q^T \Phi_j (x^*_j -x _j^{(k)} )  \label{ZZZ-error}.
\end{align}
 Denote by  $S_i = \textrm{supp} (x^*_i) $ and $\Omega_i^{(k)} =  \textrm{supp} (\widetilde{x}^{(k)}_i)=\textrm{supp} ({\cal H}_K (w_i^{(k)}))$ for $i=1, \dots, p.$ Clearly, $ |\Omega_i^{(k)}| \leq K $  for every  $ i,$ and $\sum_{i=1}^p |S_i| \leq K$ since $x^* = (x^*_1, \dots,  x^*_p) $ is $K$-sparse.
By using (\ref{ZZZ-error}) and Lemma  \ref{HHB}, we have
\begin{align}     \| \widetilde{x}^{(k)}   &   - x^*\|_{(p,\infty)}   = \|\widetilde{x}_q^{(k)}- x_q^*\|_2   = \|{\cal H}_K(w_q^{(k)})- x_q^*\|_2  \leq  \omega \| ( w_q^{(k)}- x_q^*)_{S_q\cup\Omega_q^{(k)}}\|_2 \nonumber  \\
     & \leq    \omega  \sum_{j< q} \| [\Phi_q^T \Phi_j (x^*_j - \widetilde{x}_j^{(k)})]_ {S_q\cup\Omega_q^{(k)} }\|_2   + \omega \sum_{j > q}\| [\Phi_q^T \Phi_j (x^*_j -x _j^{(k)} ]_ {S_q\cup\Omega_q^{(k)}} \|_2.    \label{Error-Ge}
\end{align}
 We now  bound the terms on the right-hand side of (\ref{Error-Ge}).  Notice that for any $j<q,$
 \begin{align*}
   \| [ \Phi_q^T \Phi_j (x_j^*- \widetilde{x}_j^{(k)})]_{ S_q\cup\Omega_q^{(k)}} \|_2 & =  \| [ (\Phi_q)_{ S_q \cup\Omega_q^{(k)}}]^T  \Phi_j  (x_j^*- \widetilde{x}_j^{(k)})  \|_2 \nonumber \\
 & =  \| [ (\Phi_q)_{S_q\cup\Omega_q^{(k)}}]^T  (\Phi_j)_{ S_j\cup\Omega_j^{(k)}}   (x_j^*- \widetilde{x}_j^{(k)})_{S_j\cup\Omega_j^{(k)} }  \|_2.
  \end{align*}
  By Lemma \ref{Lem-Ml1l2}, one has
  \begin{align}
  \| [ \Phi_q^T \Phi_j (x_j^*- \widetilde{x}_j^{(k)})]_{ S_q\cup\Omega_q^{(k)}} \|_2 & \leq \frac{1}{2}(| S_q\cup \Omega_q^{(k)}|+ | S_j\cup\Omega_j^{(k)}|) \mu(A)\| (x_j^*- \widetilde{x}_j^{(k)})_{ S_j \cup \Omega_j^{(k)}} \|_2   \nonumber  \\
  & \leq \frac{1}{2}\mu(A)(2K + | S_q|+ | S_j|) \|x^*- \widetilde{x}^{(k)} \|_{(p,\infty)},    \label{E-Bound-01}
 \end{align}
 where the last relation follows from $|\Omega_q^{(k)}| \leq K, |\Omega_j^{(k)}|\leq K$ and   the definition of  $\| \cdot\|_{(p, \infty)}$. Denote by $ \Gamma_i^{(k)} =\textrm{supp}(x^{(k)}_i). $ Since $ x^{(k)}$ is $K$-sparse, $ |\Gamma^{(k)}_i|\leq K$ for every $i=1, \dots, p.$
By the same analysis above and noting that $\textrm{supp}(x_j^*- x_j^{(k)}) \subseteq S_j \cup \Gamma_i^{(k)}$,  for any $ j> q$ one has
 \begin{align}
  \|[ \Phi_q^T \Phi_j (x_j^*- x_j^{(k)})]_{S_q\cup \Omega_q^{(k)}} \|_2
 & = \| [ (\Phi_q)_{ S_q\cup\Omega_q^{(k)}}]^T  (\Phi_j)_{ S_j\cup\Gamma_j^{(k)}}   (x_j^*- x_j^{(k)})_{S_j\cup \Gamma_j^{(k)}} \|_2   \nonumber \\
 & \leq \frac{1}{2} (| S_q \cup\Omega_q^{(k)}|+ | S_j\cup\Gamma_j^{(k)}|) \mu(A)\| (x_j^*- x_j^{(k)})_{ S_j\cup\Gamma_j^{(k)}}\|_2  \nonumber \\
 & \leq  \frac{1}{2}\mu(A) (2K + |  S_q|+ | S_j|)\|x^*- x^{(k)} \|_{(p,\infty)}.  \label{E-Bound-02}
 \end{align}
 Since $| S_q|\leq K$ and $\sum_{j<q} | S_j| \leq K,$ we see that
\begin{align}  \sum_{j< q}  2K + | S_q|+ | S_j|    & =  (2K+  |S_q| )(q-1)+   \sum_{j < q}|S_j|
\leq    (3q-2) K. \label{sum-cond-01}
\end{align}
Similarly, as  $\sum_{j>q} | S_j| \leq K,$ one has
\begin{align}  \sum_{j> q}  2K + |S_q|+ | S_j|   & = (2K+ |S_q| )(p-q)+    \sum_{j>  q} |S_j|
\leq    (3(p-q) +1) K.   \label{sum-cond-02}
\end{align}
Substituting (\ref{E-Bound-01}) and (\ref{E-Bound-02}) into (\ref{Error-Ge}) and using (\ref{sum-cond-01}) and (\ref{sum-cond-02}) yields
 \begin{align}  \|\widetilde{x}^{(k)} &  -x^*\|_{(p,\infty)}
     \leq \frac{1}{2} \omega \mu(A) \Big{[}  \sum_{j< q} (2K + | S_q|+ | S_j|) \|\widetilde{x}^{(k)}- x^*\|_{(p,\infty)}  \nonumber  \\
  &   ~~~~ + \sum_{j > q} (2K + |  S_q|+ | S_j|)  \|x^{(k)}- x^*\|_{(p,\infty)}\Big{]}   \nonumber \\
  & \leq  \frac{1}{2} \omega K \mu(A)  [ (3q-2)  \|\widetilde{x}^{(k)} - x^*\|_{(p,\infty)}  +  (3(p-q)+1)  \|  x^{(k)}-x^* \|_{(p,\infty)}] \label{ZHAO-Err-01}.
\end{align}
 As $ 1\leq q\leq p,$ we see that $ \frac{1}{2}(3q-2)\omega K \mu(A)\leq  \frac{1}{2}(3p-2)\omega K \mu(A) \leq  \frac{1}{2}(3p+1)\omega K \mu(A)<1, $ where the last inequality follows  from  (\ref{Multi-Cond})  which implies that
  \begin{equation} \label{CCKKWW}  K \mu(A) < \tau_1  := \frac{2}{\omega(1+2\omega^2(3p-2)\sqrt{p} )}<\frac{2}{\omega (1+3p)} < \frac{1}{5},\end{equation}
  where $ \tau_1<\frac{2}{\omega (1+3p)} $  follows from the fact $ 2\omega^2(3p-2)\sqrt{p}  > 3p$ for any positive integer number $ p . $
  Thus by merging terms, (\ref{ZHAO-Err-01}) can be written as
\begin{equation} \label{Final-Error-ES} \|  \widetilde{x}^{(k)} -  x^*\|_{(p,\infty)} \leq  f(q) \| x^{(k)} -  x^* \|_{(p,\infty)}.
\end{equation}
where
$$ f(q): =\frac{ \frac{1}{2} \omega K \mu(A) ( 3(p-q) +1)   } { 1- \frac{1}{2}\omega K \mu(A) (3q-2) }  \leq \rho := f(1) = \frac{ \omega K \mu(A)  (3(p-1) +1) } { 2- \omega K \mu(A)}.  $$
The inequality above follows from that $f(q)$ is decreasing over the interval $[1, p]$ provided  $\frac{3p-1}{2}\omega K \mu (A) <1 $ which is ensured since $\frac{3p+1}{2}\omega K \mu(A)<1$ due to (\ref{CCKKWW}).  $ 2-\omega K \mu(A) >0$ is also implied from (\ref{CCKKWW}).
Thus it follows from (\ref{Final-Error-ES}) that
\begin{equation} \label{Final-ER-B} \|\widetilde{x}^{(k)} -x^*  \|_{(p,\infty)} \leq  \rho \| x^{(k)}- x^* \|_{(p,\infty)}. \end{equation}
Let $ \Lambda  = {\cal L}_{\tau} (\widetilde{x} ^{(k)} )$
and $ u= {\cal H}_{\tau} (\widetilde{x}^{(k)}) .$ Clearly,   $ \textrm{supp}(u)\subseteq \Lambda.$  By Lemma \ref{HHB}, we have
$$ \|u - x^* \|_2 \leq \omega \|(\widetilde{x}^{(k)}- x^*)_{S\cup \Lambda}\|_2 \leq \omega \|\widetilde{x}^{(k)}- x^* \|_2 \leq \omega \sqrt{p}\|\widetilde{x}^{(k)}- x^* \|_{(p, \infty)}, $$
where $ S=\textrm{supp}(x^*). $  Combing this relation with (\ref{Final-ER-B}) yields
\begin{equation}\label{2702-a}  \|u-x^*\|_2 \leq \omega \rho \sqrt{p}\|x^{(k)}-x^*\|_{(p, \infty)} . \end{equation}
According to S3 of MSAA, $ \Lambda^{(k)} = {\cal L}_{\tau} (\widetilde{x}^{(k)}) \cup {\cal L}_{2K-\tau} (d^{(k)}).$
Noting that $ \textrm{supp}(u)\subseteq \Lambda \subseteq \Lambda^{(k)} $ which implies that $ u_{\overline{\Lambda^{(k)}}} =0.$
By the structure of MSAA, the $2K$-sparse vector $\widehat{x}^{(k)}$ is the solution to the least-squares problem in (\ref{MSAA03}). By using Lemma \ref{Lem-MSAA02} and noting that   $|\Lambda^{(k)}| \leq 2K < \frac{1}{\mu(A)},$ where the second inequality follows from (\ref{CCKKWW}) due to $ K\mu(A) < 2/(\omega (1+3p)) < 1/2, $   we have that
 $$ \|\widehat{x}^{(k)}-x^*\|_2 \leq \rho_1 \|(\widehat{x}^{(k)}-x^*)_{\overline{\Lambda^{(k)}}} \|_2 = \rho_1 \|(u-x^*)_{\overline{\Lambda^{(k)}}}\|_2\leq \rho_1 \|u-x^* \|_2,
$$
  where $\rho_1 = \sqrt{1+ \left(\frac{pm\mu(A)}{2(1-2K\mu(A))}\right)^2 }, $
 and  the equality above follows from $(\widehat{x}^{(k)})_{\overline{\Lambda^{(k)}}} =0 = (u)_{\overline{\Lambda^{(k)}}}. $
Denote by $ v= {\cal H}_K (\widehat{x}^{(k)}) = (\widehat{x}^{(k)} )_{S^{(k)}}, $ where $ S^{(k)} ={\cal L}_K (\widehat{x}^{(k)}) $.  It follows from Lemma \ref{HHB} and the above inequality that
 \begin{equation} \label{Error-ZYB-25} \|v-x^*\|_2\leq \omega \|(\widehat{x}^{(k)}-x^*)_{S\cup \textrm{supp}(v)}\|_2\leq \omega\rho_1 \|u-x^* \|_2. \end{equation}
 As  $ \textrm{supp} (v) \subseteq  S^{(k)} $, we see that $ (v)_{\overline{S^{(k)}}} =0.$ Notice that $|S^{(k)}|= K $ and $ x^{(k+1)} = (x_1^{(k+1)},\dots, x_p^{(k+1)})$ is the solution to (\ref{MSAA04}), by Lemma \ref{Lem-MSAA02} again, one has
$$ \|x^{(k+1)} - x^* \|_2 \leq \rho_2  \|(x^{(k+1)} - x^*)_{\overline{S^{(k)}}}\|_2=  \rho_2 \|(v - x^*)_{\overline{S^{(k)}}}\|_2 \leq \omega \rho_1\rho_2  \|u-x^* \|_2, $$
where  $\rho_2= \sqrt{1+ \left(\frac{pm\mu(A)}{2(1-K\mu(A))}\right)^2 }, $  the equality follows from $ (x^{(k+1)})_{\overline{S^{(k)}}} =0 =(v)_{\overline{S^{(k)}}}$, and the last inequality from (\ref{Error-ZYB-25}).
Merging (\ref{2702-a}) and the inequality above yields
\begin{equation}\label{F-error} \|x^{(k+1)} - x^*\|_2 \leq \omega^2 \rho \rho_1\rho_2 \sqrt{p} \|x^{(k)}-x^*\|_{(p,\infty)}=  \eta \|x^{(k)}-x^*\|_{(p,\infty)},
\end{equation}
where $ \eta:=   \omega^2 \rho \rho_1\rho_2\sqrt{p}. $ That is,  $$
\eta  =  \frac{\omega^3 K \mu(A)  (3p-2)\sqrt{p} } { 2- \omega K \mu(A)} \sqrt{ \left[1+ \left(\frac{pm\mu(A)}{2(1-2K\mu(A))}\right)^2\right] \left[1+ \left(\frac{pm\mu(A)}{2(1-K\mu(A))}\right)^2\right] }.
$$ Thus from (\ref{F-error}), to show $ x^{(k)} \to  x^*$, it is sufficient to
 verify that $ \eta <1$ under (\ref{Multi-Cond}). Since  $ 1-K \mu(A) > 1-2K\mu(A) >0 $ which is ensured by (\ref{Multi-Cond}),  $\eta $ is  bounded as
 $$ \eta \leq  \frac{\omega^3 K \mu(A)  (3p-2)\sqrt{p}} { 2- \omega K \mu(A)} \left[1+ \left(\frac{pm\mu(A)}{2(1-2K\mu(A))}\right)^2\right] . $$
 If $ \frac{pm\mu(A)}{2(1-2K\mu(A))} \leq 1,$ then $  \eta \leq   \frac{2\omega^3 K \mu(A)  (3p-2)\sqrt{p}  } { 2- \omega K \mu(A)}   <1,  $ where the second inequality follows from $ K\mu(A) < \tau_1$ which is  (\ref{CCKKWW}).
 If $ \frac{pm\mu(A)}{2(1-3K\mu(A))} > 1,$ we have
 \begin{equation} \label{eta-b} \eta  \leq \frac{2\omega^3 K \mu(A)  (3p-2)\sqrt{p}} { 2- \omega K \mu(A)}\left(\frac{pm\mu(A)}{2(1-2K\mu(A))}\right)^2  <  \frac{\omega^3 K \mu(A)^3  (3p-2)\sqrt{p} (pm)^2 } { 2(2- \omega \tau_1)(1-2\tau_1)^2} , \end{equation}
 which follows from $ 2- \omega K \mu(A) \geq  2- \omega \tau_1 >0 $ and $ 1-2K\mu(A) \geq 1-2\tau_1 >0 $ due to $ K\mu(A) < \tau_1$ and $ \tau_1 <1/5 < 2/\omega$ (by \ref{CCKKWW})). Define   $\tau_2:= \frac{2(2- \omega \tau_1)(1-2\tau_1)^2}{\omega^3   (3p-2)p^2\sqrt{p} }  . $ Then   (\ref{eta-b}) can be written as $ \eta < K\mu(A)^3m^2/\tau_2 $ which is strictly less than 1 under the condition in (\ref{Multi-Cond}).  Thus $\eta <1$ is guaranteed by  (\ref{Multi-Cond}).
 From (\ref{F-error}), we deduce that  $\{x^{(k)}\}_{k\geq 1} $  converges to $ x^*$. Since the right-hand side of (\ref {Multi-Cond}) is lower than $ \frac{1}{2} (1+\frac{1}{\mu(A)}),$  $ x^*$ is the unique sparsest solution to the linear system by Lemma \ref{Lem-M-classic}.  ~~~ $ \Box $

\begin{remark} (i) Theorems \ref{Thm-Mu-Co} and \ref{Main-Thm-Mu-Co} are established for TSAA and MSAA, respectively, for the first time. Conditions (\ref{Mu-Condition}) and (\ref{Multi-Cond}) can be satisfied when $\mu(A)$ is small and/or the sparsity level of $x^*$ is low. For example, let $U$ denote the right-hand side of (\ref{Mu-Condition}). Clearly, $U > 1$ when $\mu(A)$ is sufficiently small. In this case, let $x^*$ be a $K$-sparse vector with $K := \lfloor U \rfloor,$ and set $y := Ax^*$ as the measurements. Then the system $Ax = y$ will satisfy the condition in (\ref{Mu-Condition}). \smallskip (ii) Although this paper focuses on problems involving concatenated square orthogonal matrices (where all blocks $\Phi_i, i = 1, \dots, p,$ have the same number of columns), the algorithms presented here can be extended to a more general setting where the splitting blocks are not necessarily square. The extended algorithm and its convergence analysis would differ significantly from those in this paper. A RIP-based convergence analysis might be more convenient for the extended algorithms, which is a worthwhile future work.   
\end{remark}

\begin{remark} TSAA and MSAA are shown to be convergent under (\ref{Mu-Condition}) and (\ref {Multi-Cond}), respectively.  Both assumptions pertain to the range of sparsity level, i.e., $ K < K^*$, where the bound $K^*$ depends on $\mu(A). $  A larger bound $K^*$ implies that a wider range of problems satisfy the assumption. Moreover, Theorems  \ref{Thm-Mu-Co} and \ref{Main-Thm-Mu-Co}  both characterize  the decay speed of the error sequence $\{\|x^{(k)}-x^*\|_2\}$ in terms of the decay  ratio $ \rho$ and $\eta$, respectively.  Clearly,  a smaller decay ratio indicates faster algorithm convergence.  Theorem \ref{Thm-Mu-Co}, established through a separate analysis, is not a special case of Theorem \ref{Main-Thm-Mu-Co} .  It cannot be derived from the latter by merely setting $p=2.$  In fact,  Theorem \ref{Thm-Mu-Co} is more profound than Theorem \ref{Main-Thm-Mu-Co}  for $ p=2$ in two aspects: (\ref{Mu-Condition}) is more relaxed than (\ref {Multi-Cond}), and the convergence speed claimed in Theorem \ref{Main-Thm-Mu-Co}  is slower than that in Theorem \ref{Thm-Mu-Co}.  To elaborate, let us denote $\tau_1$ and $ \tau_2$ in Theorem \ref{Main-Thm-Mu-Co}   as $\tau_1 (p)$ and $ \tau_2 (p) ,$ respectively, since they depend on $p. $  Note that $ \tau_1(p) $ and $\tau_2(p) $   are strictly decreasing with respect to $p \geq 2.$ Thus it is straightforward to verify that  $$\max_{p\geq 2}\tau_1(p)=\tau_1(2)<\frac{c}{2}, ~ \max_{p\geq 2}\tau_2(p)=\tau_2(2) <\frac{c}{2}(1-c)^2,$$   where $ c= \frac{\sqrt{2}}{3\omega^3}$ with $\omega = \frac{\sqrt{5}+1}{2}$. 
This implies that for any $ p\geq 2,$  \begin{equation*}
    \min\left\{\tau_1(p),~\tau_2(p)\left(\frac{1}{m\mu(A)}\right)^2\right\} <   \min\left\{\frac{c}{2}, ~ \frac{c(1-c)^2}{2} \left( \frac{1}{m\mu (A)}\right)^2\right\} .
 \end{equation*} Consequently,  (\ref{Mu-Condition}) in Theorem \ref{Thm-Mu-Co} is more relaxed than (\ref {Multi-Cond}) in Theorem \ref{Main-Thm-Mu-Co}  for any $ p\geq 2.$  Furthermore, it can be verified that under (\ref {Multi-Cond}), the decay ratio $\rho$ in Theorem \ref{Thm-Mu-Co} is strictly smaller than $\eta$ in Theorem \ref{Main-Thm-Mu-Co}. Since $ \eta$ depends on $ p$, we denote it as $ \eta(p) $ which is strictly increasing in $p.$  It is not difficult to verify that under (\ref {Multi-Cond}),  $\rho< \frac{3}{4\sqrt{2}}\eta(2) < \frac{3}{4\sqrt{2}}\eta(p)$ for any integer $ p>2. $ This means  the convergence speed claimed in Theorem \ref{Main-Thm-Mu-Co}  is slower than that in Theorem \ref{Thm-Mu-Co}.
  The analysis for two-block case can be conducted more deeply, as only two blocks need to coordinate. For this case, when the partition $y_a$ for one block  is given, the partition $y_b = y - y_a$ for the other block is immediately determined. However, the analysis for the case  $ p >2 $ encounters much more challenges due to the increased uncertainty and complexity associated with multiple-block coordination and  observation partition between blocks.
\end{remark}

\section{Numerical experiments} \label{Num-exp}
 
 The performance of our algorithms on synthetic data is demonstrated and compared with existing ones, including IHT, HTP, SP, CoSaMP, OMP, $\ell_1$-minimization, and FISTA with a fixed stepsize. Additionally, the performance of TSAA on real magnetic resonance image (MRI) reconstruction is demonstrated. In our experiments, $\ell_1$-minimization ($\ell_1$-min) is solved using CVX with the \emph{`Mosek'} solver \cite{GB17}. The steplength in IHT and HTP is set to 1, as the columns of $A$ are normalized. The number of iterations for OMP is set to the sparsity level $K$ of the vector to be recovered. We use IT to denote the total number of iterations performed by an algorithm. Initial simulations indicate that the success rate of our algorithms for locating the sparse solution of linear systems is insensitive to the choice of initial points. Thus, we set $x^{(0)}=0$ (together with $y^{(0)}_i = y/p$ for $i=2,\dots, p$) as the default initial point. The value of $\tau$ ($K \leq \tau \leq 2K$) reflects how much gradient information at the current iterate is used to reduce the error metric via the projection in Step 3 of TSAA and MSAA. Initial simulations also indicate that using part of the gradient information can enhance the performance of our algorithms. Thus we set $\tau = K$ as the default value for this parameter.

\subsection{Performance on synthetic data}\label{sec-rand}
The Matlab codes \emph{`sprandn(n,1,d)'} and \emph{`orth(randn(m))'} are used to generate  the $K$-sparse vector $x^*\in\mathbb{R}^{ n}$ and orthogonal matrices $\Phi_i \in\mathbb{R}^{m\times m}(i=1,\ldots,p),$   where $n=m*p$ and $d=K/n\in(0,1).$ Such vectors have at most $K$ normally distributed nonzero elements,  and $\Phi_i $'s are the orthogonalized random  Gaussian matrices. The performance of  algorithms can be evaluated by their success rate for recovering sparse vectors via accurate or inaccurate measurements. In our experiments, the recovery criterion is set as
\begin{equation}\label{rel_er}
 \|x^{(k)}-x^*\|_2/\|x^*\|_2\leq 10^{-4}.
\end{equation}
The recovery of $ x^*$ is said to be successful if the vector $ x^{(k)} $ generated by the algorithm satisfies (\ref{rel_er}).  The matrix
$A=[\Phi_1,\ldots,\Phi_p]\in\mathbb{R}^{m\times n}$ is of size $m=1000$ and $ n=pm$ with $p=2$ or 5. When the sparsity level $K$ is pretty low, all algorithms mentioned in this section can successfully recover the sparse vectors. Thus we only demonstrate the performance of algorithms on the vectors with relatively high sparsity levels, i.e.,  $  K_0 \leq K \leq  K_{max},$ where  $K_0=20$ and $ K_{max} > m/2.$   For every given sparsity level $K= K_0+5i$ for $ i =0,1, \dots,  \lfloor\frac{K_{max}-K_0}{5} \rfloor,$ we use 100 random examples of  $(A, x^*)$  to evaluate the success rate of the algorithms.

\subsubsection{Solving problems with just a few iterations}

\begin{figure}[htbp]
     \centering
 \subfloat[TSAA: IT=3, 4, 5, 6, 7]{
\begin{minipage}[t]{0.45\linewidth}
  \includegraphics[interpolate=true,width=\textwidth,height=0.77\textwidth]{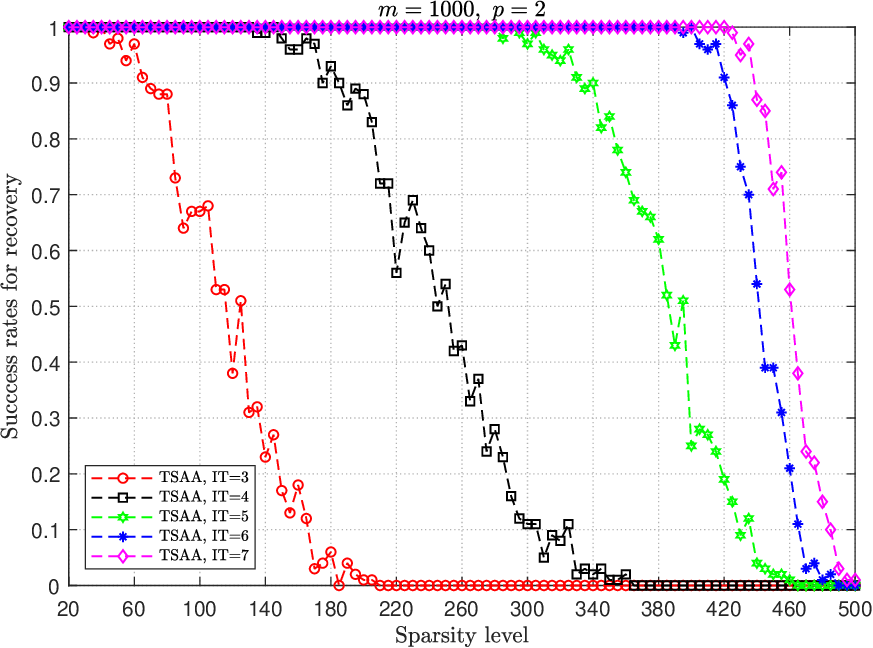}
   \end{minipage}}\hspace{0.2cm}
   \subfloat[Algorithms with IT=7]{
\begin{minipage}[t]{0.45\linewidth}
  \includegraphics[interpolate=true,width=\textwidth,height=0.77\textwidth]{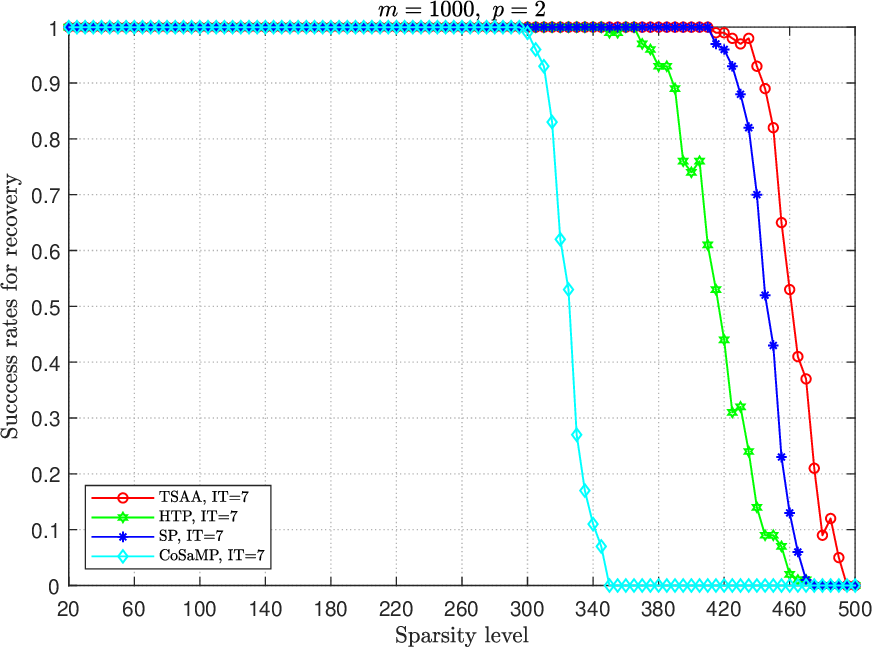}
   \end{minipage}}
    \caption{Success rates for recovery with just a few iterations.}
  \label{diff-IT}
   \end{figure}

The first observation is that our algorithms require only a few iterations to find the sparsest solution of a wide range of underdetermined linear systems. In signal-recovery terms, our algorithms can recover a broad range of sparse signals using just a few iterations. Here, we demonstrate the results for TSAA in noiseless situations, while similar results are also observed in noisy settings.The results for IT = 3, \dots, 7 are summarized in Fig. \ref{diff-IT}(a), which shows that all $K$-sparse vectors with $K \leq 0.4$ m are exactly recovered by TSAA in just a few iterations. The range of problems that TSAA can solve broadens with each additional iteration.We also observe that TSAA can solve a wider range of problems than several mainstream iterative methods (HTP, SP, CoSaMP) when using the same number of their first few iterations. Their comparison is given in Fig. \ref{diff-IT}(b), where all algorithms are run for only 7 iterations.

\subsubsection{Comparison of overall success rates and runtime}
In our experiments, the default maximum number of iterations for  TSAA, MSAA, HTP, SP and CoSaMP is set to be 100; the parameter $\lambda$ and the Lipschitz constant $L$ in FISTA \cite{BT09} are set to be  $\lambda =4\times 10^{-5} $ and  $L=\lambda_ {max}(A^TA)$ (the largest eigenvalue of $A^TA$); IHT and FISTA are allowed to perform up to 3000 iterations due to their   slow convergence. From a  practical point of view, it is more sensible to evaluate the performance of the algorithms in noisy situations. Thus we take the  measurements   $ y: = Ax^*+\epsilon h, $ where $h\in\mathbb{R}^{ m}$ is  a standard Gaussian noise vector and   $\epsilon> 0$ is the noise level. We stop the algorithm when $\|x^{(k)}-x^{(k-1)}\|_2/\|x^{(k)}\|_2\leq 10^{-8} $ or the prescribed maximum number of iterations whichever is first reached, and then adopt \eqref{rel_er}  to decide whether the recovery of $ x^*$ is successful.

\begin{figure}[htbp]
   \centering
 \subfloat[ $p=2$ and $ \epsilon=5\times 10^{-5}$]{
\begin{minipage}[t]{0.45\linewidth}
  \includegraphics[interpolate=true,width=\textwidth,height=0.77\textwidth]{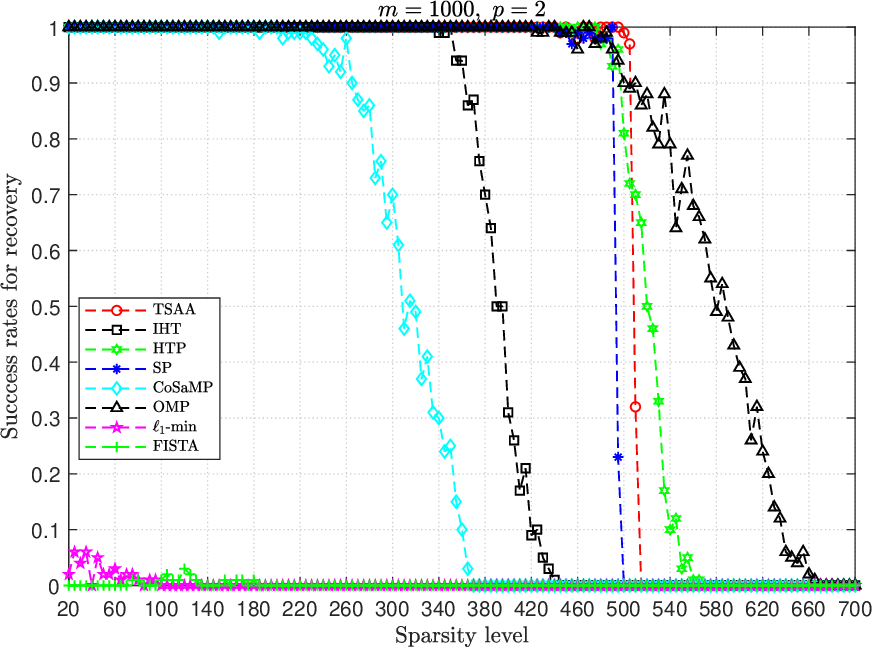}
   \end{minipage}
   }\hspace{0.2cm}
   \subfloat[ $p=5$ and $\epsilon=3\times 10^{-5}$]{
\begin{minipage}[t]{0.45\linewidth}
  \includegraphics[interpolate=true,width=\textwidth,height=0.77\textwidth]{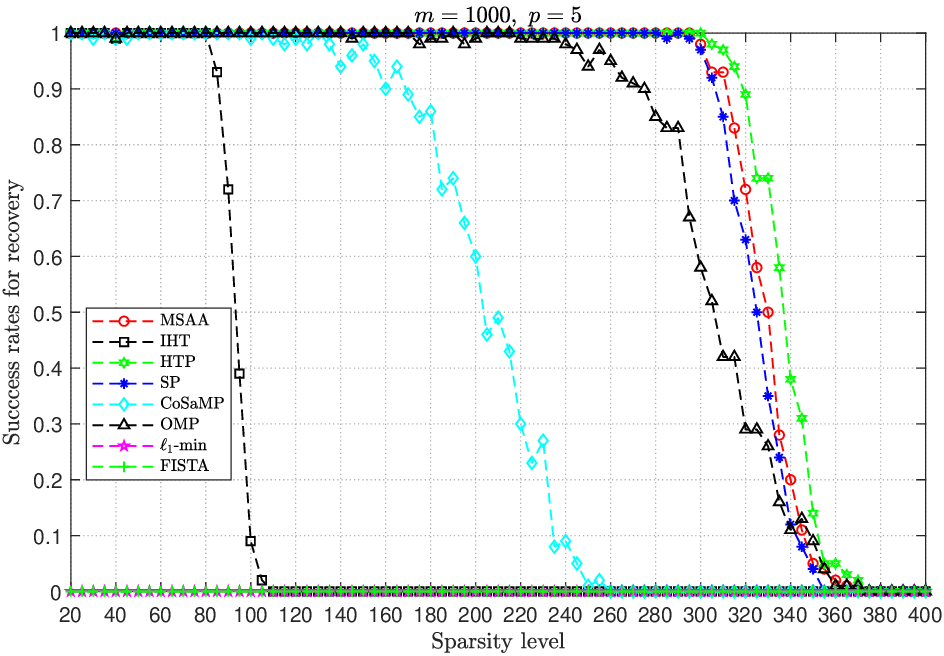}
   \end{minipage}}
   \caption{Comparison of success rates  for vector recovery. (a) Two-block case. (b) Five-block case.}
  \label{suc}
   \end{figure}

In this experiment, the success rate of each algorithm for a given $K$ is determined using 100 random problem instances. The performance of TSAA and MSAA, compared to several existing algorithms, is summarized in Fig. \ref{suc}. The result in Fig. \ref{suc}(a) is obtained under the noise level $\epsilon=5\times 10^{-5},$ while Fig. \ref{suc}(b) is obtained under $\epsilon=3\times 10^{-5}.$ The results indicate that $\ell_1$-min and FISTA are severely affected by the given noise levels. From Fig. \ref{suc}(a), the success rates of IHT and CoSaMP drop to 35\% when $K/m$ is close to 0.4, but TSAA continues to succeed even as $K/m$ approaches 0.5. This demonstrates that TSAA significantly outperforms IHT, CoSaMP, $\ell_1$-min, and FISTA, and is comparable to HTP, SP, and OMP. Fig. \ref{suc}(b) shows that MSAA with $p=5$ remains very robust, with an overall performance better than OMP and comparable to HTP and SP. This experiment indicates that our algorithms are stable, as their performance is not very sensitive to changes in noise levels or the measurement rate $m/n$ which, however, severely affect the performance of IHT, $ \ell_1$-min, FISTA and OMP.

\begin{figure}[htbp]
     \centering
 \subfloat[ $p=2$ and $ \epsilon=5\times 10^{-5}$]{
\begin{minipage}[t]{0.45\linewidth}
 \includegraphics[interpolate=true,width=\textwidth,height=0.77\textwidth]{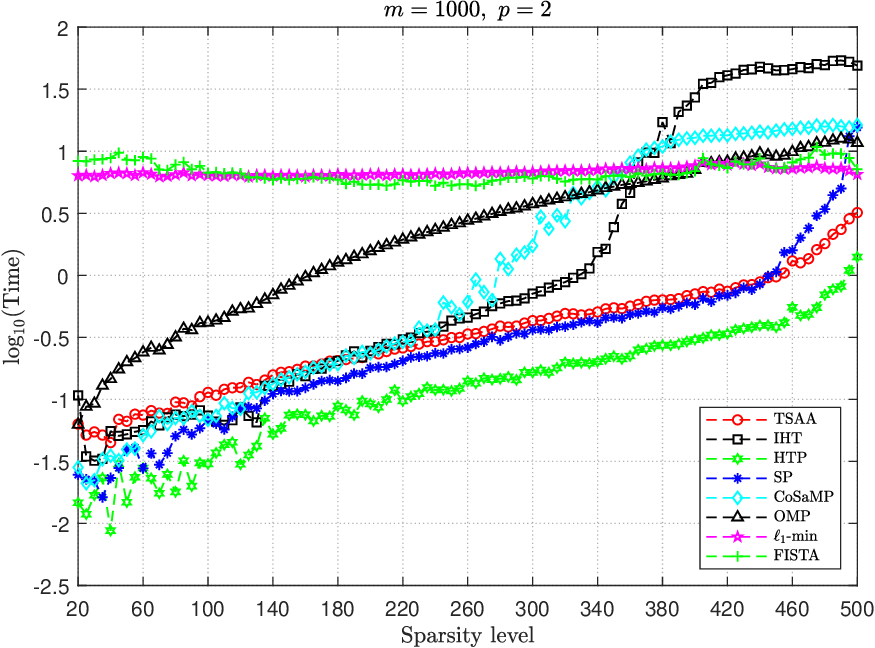}
  \end{minipage}
   }\hspace{0.2cm}
   \subfloat[ $p=5$ and $  \epsilon=3\times 10^{-5}$]{
\begin{minipage}[t]{0.45\linewidth}
 \includegraphics[interpolate=true,width=\textwidth,height=0.77\textwidth]{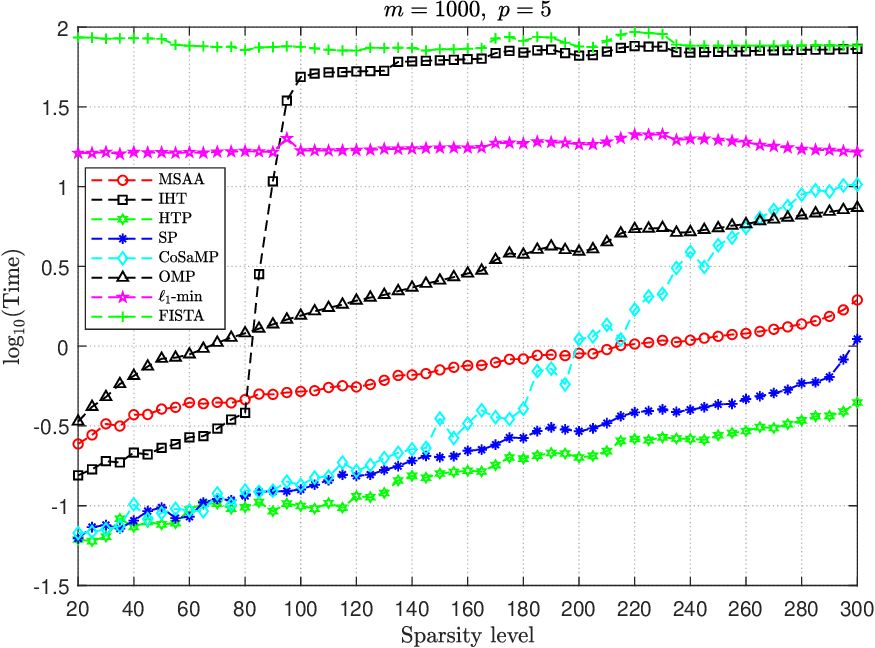}
  \end{minipage} }
   \caption{Comparison of CPU time (in seconds) spent by algorithms  to achieve recovery success. (a) Two-block case. (b) Five-block case.}
 \label{time}
  \end{figure}

The CPU time spent in these experiments is given in Fig. \ref{time}. Fig. \ref{time}(a) shows that TSAA consumes remarkably less time than FISTA, $\ell_1$-min, and OMP to achieve the same recovery criterion (\ref{rel_er}), but spends slightly more time than HTP. The runtime is also lower than that of SP, CoSaMP, and IHT as $K$ increases. Similar results are observed for MSAA with $p=5,$ as shown in Fig. \ref{time} (b).  

\subsection{Reconstruction of  MRI images}\label{sec-image}

Two brain MRI images, {\it Image 1} and {\it Image 2}, of size $192 \times 174$ are reconstructed using TSAA and several existing methods for comparison. We represent the original MRI image by concatenating its columns into a vector $x^* \in \mathbb{R}^n$ with $n = 33408.$ The coefficient vector $c^* \in \mathbb{R}^n$ of $x^*$ under the discrete wavelet transform $\mathbf{W}(\cdot)$ is used, where $\mathbf{W}$ is based on the wavelet \emph{`db1'} with 6 levels and periodic extension mode. Specifically, $c^* = \mathbf{W}(x^*)$, which is often compressible. We obtain the accurate measurements $y := A c^* $, where $A \in \mathbb{R}^{m \times n}$ ($m = n/2$) is constructed by concatenating two orthogonal matrices. Our goal is to generate a $K$-sparse vector $\hat{c} \in \mathbb{R}^n,$ which is the best $K$-term approximation to $c^*.$ We set $K = m/3$ for TSAA, IHT, HTP, SP, and CoSaMP. The reconstructed image $\hat{x} \in \mathbb{R}^n$ is then obtained by $\hat{x} = \mathbf{W}^{-1}(\hat{c}),$ where $\mathbf{W}^{-1}$ is the inverse of $\mathbf{W}.$ The quality of the reconstructed image is assessed using the peak signal-to-noise ratio (PSNR):  
$
\textrm{PSNR} =10\cdot \log_{10}(255^2/\textrm{MSE}),
$
where MSE denotes the mean-squared error between the original and reconstructed images.

\begin{figure}[htbp]
   \centering
 \subfloat[Image 1(Original)]{\label{fig:a}\includegraphics[width=0.22\textwidth,height=0.22\textwidth]{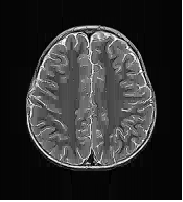}} \hspace{0.1cm}
 \subfloat[IT=1]{\label{fig:b}\includegraphics[width=0.22\textwidth,height=0.22\textwidth]{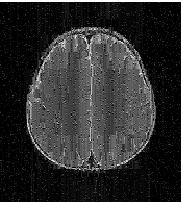}}  \hspace{0.1cm}
  \subfloat[IT=4]{\label{fig:c}\includegraphics[width=0.22\textwidth,height=0.22\textwidth]{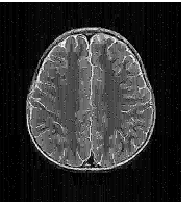}}  \hspace{0.1cm}
  \subfloat[IT=9]{\label{fig:d}\includegraphics[width=0.22\textwidth,height=0.22\textwidth]{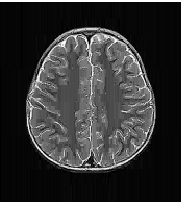}}
  \\
  \subfloat[Image 2(Original)]{\label{fig:e}\includegraphics[width=0.22\textwidth,height=0.22\textwidth]{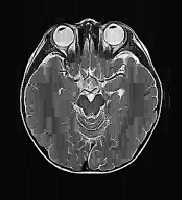}}\hspace{0.1cm}
  \subfloat[IT=1]{\label{fig:f}\includegraphics[width=0.22\textwidth,height=0.22\textwidth]{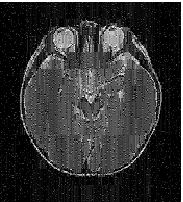}} \hspace{0.1cm}
  \subfloat[IT=4]{\label{fig:g}\includegraphics[width=0.22\textwidth,height=0.22\textwidth]{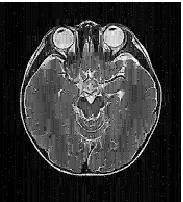}} \hspace{0.1cm}
  \subfloat[IT=9]{\label{fig:h}\includegraphics[width=0.22\textwidth,height=0.22\textwidth]{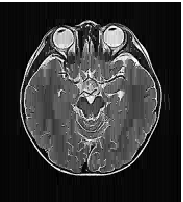}}
     \caption{Reconstructed images by performing TSAA just a few iterations: IT=1, 4, 9. (Original Image 1 and Image 2  courtesy of the Shandong Zibo Central Hospital.)}
     \label{TSAA_Rec_img}
\end{figure}

\begin{figure}[htbp]
     \centering
 \subfloat[PSNR vs IT]{
\begin{minipage}[t]{0.45\linewidth}
  \includegraphics[interpolate=true,width=\textwidth,height=0.77\textwidth]{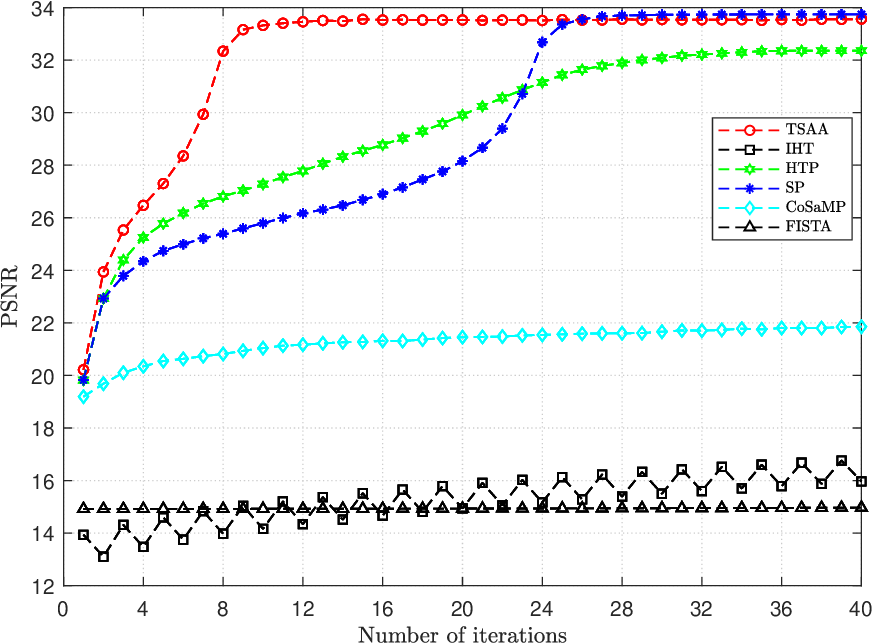}
  \end{minipage}
  }\hspace{0.2cm}
   \subfloat[CPU time vs PSNR]{
\begin{minipage}[t]{0.45\linewidth}
 \includegraphics[interpolate=true,width=\textwidth,height=0.77\textwidth]{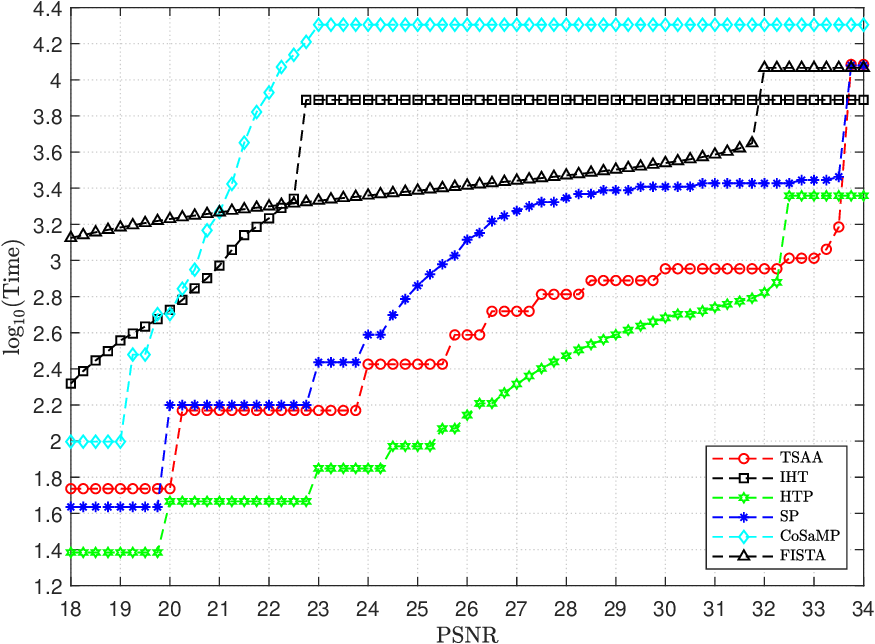}
  \end{minipage} }\hspace{0.2cm}
    \subfloat[PSNR vs IT]{
\begin{minipage}[t]{0.45\linewidth}
  \includegraphics[interpolate=true,width=\textwidth,height=0.77\textwidth]{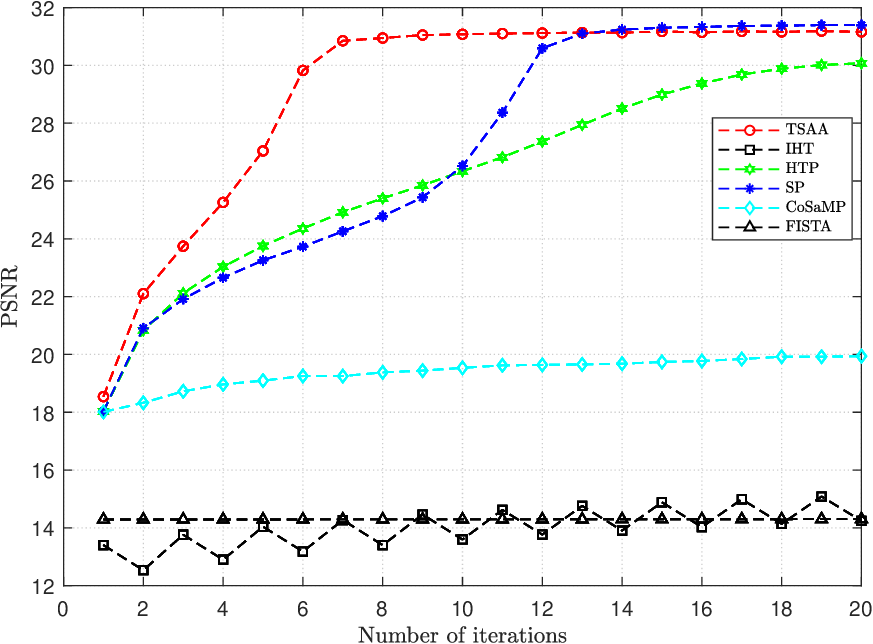}
   \end{minipage}
   }\hspace{0.2cm}
   \subfloat[CPU time vs PSNR]{
\begin{minipage}[t]{0.45\linewidth}
 \includegraphics[interpolate=true,width=\textwidth,height=0.77\textwidth]{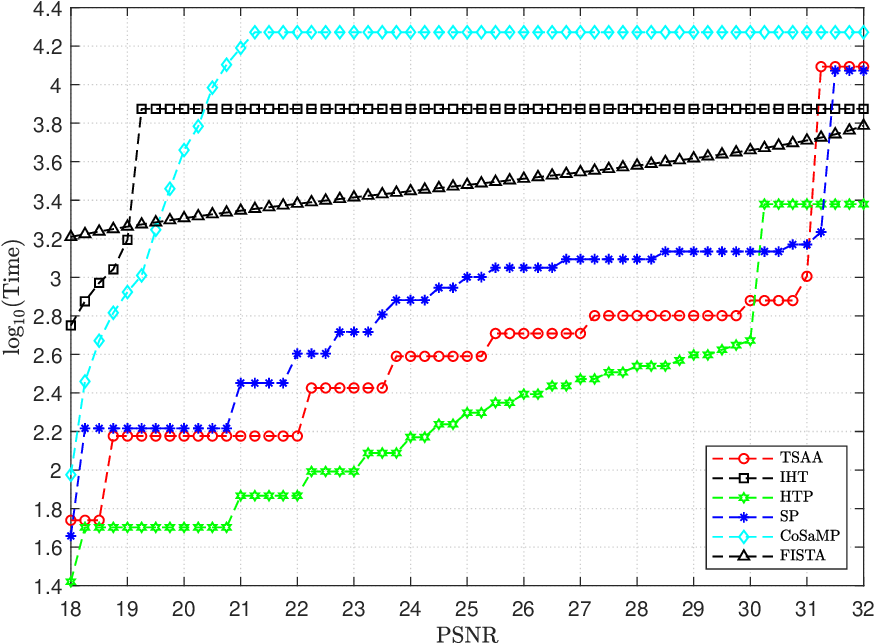}
  \end{minipage} }
  \caption{(a) and (c): Comparison of  PSNR  (in dB)  vs the number of iterations performed. (b) and (d): Comparison of  CPU time  (in seconds) vs PSNR (in dB). The first row corresponds to Image 1, and the second to Image 2.}
  \label{psnr_time}
   \end{figure}

The original MRI images and those reconstructed by TSAA are shown in Fig. \ref{TSAA_Rec_img}. The first column displays the original images, while the other three columns present the reconstructed images with IT = 1, 4, and 9 iterations, respectively. Clearly, the quality of the reconstructed images is significantly enhanced with each additional few iterations of TSAA. As expected, TSAA achieves high-quality reconstruction with very few iterations. Furthermore, it is interesting to observe how the PSNR improves as IT increases. The results are displayed in Figs. \ref{psnr_time}(a) and \ref{psnr_time}(c). From Fig. \ref{psnr_time}(a), the PSNR of TSAA rapidly reaches its peak values within only 10 iterations, surpassing other algorithms by at least 6 dB. For IT $\geq$ 24, the PSNRs of SP, HTP, and CoSaMP gradually approach their maximum values. Even when IT is increased to 40, none of the other algorithms exceed the PSNR of TSAA. Additionally, the PSNRs of IHT and FISTA improve very slowly with increasing IT, requiring a large number of iterations to achieve a certain level of reconstruction quality. Similar results are shown for Image 2 in Fig. \ref{psnr_time}(c).

Finally, let us compare the time required by the algorithms to achieve a prescribed PSNR. Specifically, we aim to determine how long each algorithm takes to reach a desired PSNR level. The results for Image 1 are shown in Fig. \ref{psnr_time}(b), and those for Image 2 are in Fig. \ref{psnr_time}(d). For Image 1, the PSNR values range from 18 to 34, while for Image 2, they range from 18 to 32, both with a step size of 0.25. In both Fig. \ref{psnr_time}(b) and Fig. \ref{psnr_time}(d), the fixed values on the far right of each curve indicate that the algorithm reached the prescribed maximum number of iterations without achieving the target PSNR. From Fig. \ref{psnr_time}(b), it is evident that TSAA can reconstruct Image 1 up to approximately PSNR = 33.5 within 100 iterations. Only TSAA and SP can achieve PSNR values in the range [32.5, 33.5], with TSAA being significantly faster than SP. For PSNR $\leq$ 32.5, TSAA is much faster than the other existing methods except for HTP, which is faster than TSAA when PSNR is below 32.5. However, HTP cannot achieve PSNR values above this threshold. CoSaMP and IHT can only reconstruct Image 1 up to PSNR = 23, even when IHT is allowed to perform 3000 iterations. The results for Image 2 in Fig. \ref{psnr_time}(d) are similar to those for Image 1, except that FISTA can achieve PSNR = 32 for this image, which is better than the other algorithms. However, FISTA is much slower than TSAA in reaching the PSNR up to 31.

\section{Conclusions}
 The splitting alternating algorithms for finding sparse solutions of underdetermined linear systems have been proposed. These algorithms are specifically tailored for large-scale linear systems involving concatenated orthogonal matrices. Their global convergence has been established under a mutual-coherence-type condition. Numerical results demonstrate that the proposed algorithms can successfully identify the sparse solution for a wide range of linear systems, often within just a few iterations.

 \section*{Acknowledgement} We would like to thank anonymous reviewers and associate editor for their valuable comments and suggestions, which helped us improve the quality of the manuscript.

 \end{document}